\documentclass[preprints,article,accept,moreauthors,pdftex]{Definitions/mdpi}

\firstpage{1} 
\pubvolume{1}
\issuenum{1}
\articlenumber{0}
\pubyear{2021}
\copyrightyear{2020}
\datereceived{} 
\dateaccepted{} 
\datepublished{} 
\hreflink{https://doi.org/} 
\pdfoutput=1

\usepackage{siunitx}
\usepackage{mathtools}
\usepackage{lineno}

\newcommand*\patchAmsMathEnvironmentForLineno[1]{%
  \expandafter\let\csname old#1\expandafter\endcsname\csname #1\endcsname
  \expandafter\let\csname oldend#1\expandafter\endcsname\csname end#1\endcsname
  \renewenvironment{#1}%
     {\linenomath\csname old#1\endcsname}%
     {\csname oldend#1\endcsname\endlinenomath}}%
\newcommand*\patchBothAmsMathEnvironmentsForLineno[1]{%
  \patchAmsMathEnvironmentForLineno{#1}%
  \patchAmsMathEnvironmentForLineno{#1*}}%
\AtBeginDocument{%
\patchBothAmsMathEnvironmentsForLineno{equation}%
\patchBothAmsMathEnvironmentsForLineno{align}%
\patchBothAmsMathEnvironmentsForLineno{flalign}%
\patchBothAmsMathEnvironmentsForLineno{alignat}%
\patchBothAmsMathEnvironmentsForLineno{gather}%
\patchBothAmsMathEnvironmentsForLineno{multline}%
}

\def\R{\mathbb{R}}
\def\L{\mathcal{L}}

\DeclareMathOperator*{\argmax}{arg\,max}
\DeclarePairedDelimiter\abs{\lvert}{\rvert}
\newcommand \dd[1]  { \,\textrm d{#1}                       }   

\Title{Conditional Invertible Neural Networks for Medical Imaging}

\TitleCitation{Conditional Invertible Neural Networks for Medical Imaging}

\Author{Alexander Denker$^{1, *}$\orcidA{}, Maximilian Schmidt$^{1}$\orcidB{}, Johannes Leuschner$^{1}$\orcidC{} and Peter Maass$^{1}$\orcidD{}
}

\AuthorNames{Alexander Denker, Maximilian Schmidt, Johannes Leuschner and Peter Maass}

\AuthorCitation{Denker, A.; Schmidt, M.; Leuschner, J.}

\address[1]{%
$^{1}$ \quad Center for Industrial Mathematics, University of Bremen, Bibliothekstr. 5, 28359 Bremen, Germany}

\corres{Correspondence: adenker@uni-bremen.de}

\abstract{Over the last years, deep learning methods have become an increasingly popular choice to solve tasks from the field of inverse problems. Many of these new data-driven methods have produced impressive results, although most only give point estimates for the reconstruction. However, especially in the analysis of ill-posed inverse problems, the study of uncertainties is essential. In our work, we apply generative flow-based models based on invertible neural networks to two challenging medical imaging tasks, i.e.\ low-dose computed tomography and accelerated medical resonance imaging. We test different architectures of invertible neural networks and provide extensive ablation studies. In most applications, a standard Gaussian is used as the base distribution for a flow-based model. Our results show that the choice of a radial distribution can improve the quality of reconstructions.}


\begin{document}


\section{Introduction}
The image reconstruction task arising in computed tomography (CT) or medical resonance imaging (MRI) can be formulated as an inverse problem. A forward operator $\mathcal{A}: X \rightarrow Y$ maps the image $x^\dagger$ to (noisy) measurements 
\begin{align}
    y^\delta = \mathcal{A}x^\dagger + \epsilon,
\end{align}
where $\epsilon \in Y$ describes the noise. Research in inverse problems has mainly focused on developing algorithms for obtaining stable reconstructions of the true image $x^\dagger$ in the presence of noise. In recent years, data-driven methods have been increasingly used in research and applications to solve inverse problems \citep{arridge2019}. The choice of methods ranges from post-processing approaches \citep{jin2017deep}, unrolling iterative schemes as neural network layers \citep{adler2018pd, adler2017solving}, and learned regularization terms \cite{lunz2018adversarial} to complete learning of an inversion model from data \citep{zhu2018automap}. However, many data-driven methods only give a point estimate of the solution as output. But especially for ill-posed inverse problems, an estimation of the uncertainties is essential.
In order to incorporate uncertainties arising in the inversion process, the reconstruction process can be interpreted in a statistical way as a \textit{quest for information} \cite{tarantola1982inverse, kaipio2005statistical}. Instead of approximating a single point estimate, we are interested in the entire conditional distribution $p(x|y^\delta)$ of the image given the noisy measurement data. Traditionally methods like Markov chain Monte Carlo \citep{martin2012stochastic} or approximate Bayesian computation \citep{sunnaaker2013approximate} have been used to estimate the unknown conditional distribution. However, these methods are often computationally expensive and unfeasible for large-scale imaging problems. 
A new approach is the application of deep generative models for this task. In general, the goal of a deep generative model is to learn a surrogate model for the unknown distribution based on samples. Well-known approaches from the field of generative networks are variational autoencoders (VAE) \citep{rezende2014stochastic, kingma2014autoencoding} and generative adversarial networks (GAN) \citep{goodfellow2014generative}. Recently, flow-based generative models \citep{tabak2013family} were introduced, which use an invertible transformation to learn a continuous probability density. One of the advantages is that flow-based models allow exact likelihood computation, thus allowing for maximum likelihood training. 

\subsection{Related Work}
A variety of neural network methods have been proposed to analyze inverse problems \citep{arridge2019}. We are especially interested in methods, that can estimate the uncertainties arising in the inversion process. Several approaches have been developed in the past, e.g.\ Bayesian neural network can be combined with deep learning models \citep{barbano2021quantifying}, or conditional GANs can be used to learn the unknown posterior density implicitly \citep{adler2018deep}. Recently, flow-based models have been used to learn a surrogate model for the unknown posterior. These flow-based models are often implemented using invertible neural networks. They have been used to predict oxygen saturation in tumors \citep{Ardizzone2019b}, image colorization \citep{Ardizzone2019a}, day-to-night translation \citep{ardizzone2021conditional}, or the identification of the permeability field of an oil reservoir \citep{ANANTHAPADMANABHA2021110194}. There is also the first application for computed tomography \cite{Denker2020, leuschner2021quantitative}. Our work builds on the concept of conditional invertible neural networks (cINNs) as introduced in \citep{Ardizzone2019a}, but our focus lies on medical image reconstruction.     

\subsection{Contributions}
Prior work of cINNs for inverse problems deals mainly with image-to-image problems \citep{Ardizzone2019a, ardizzone2021conditional} or lower dimensional applications \citep{Ardizzone2019b}. These cINNs are implemented using two components: an invertible neural network, used for the normalizing flows, and a conditioning network, used to extract features from the conditional input. This conditioning network does not have to be invertible and is often implemented as a convolutional neural network (CNN). In our work, we expand these concepts to inverse problems in medical imaging, where the topology of the measurement space and the image space differ significantly. In CT reconstruction, the measurements are line integrals over the image domain. In MR imaging, the measurements can be interpreted in the frequency domain. This creates an additional challenge as CNNs are built to take advantage of local relationships and often fail when there are global relationships in the measurements. We address this problem by integrating a traditional reconstruction operator into the conditioning network of the cINN. For the problem of CT reconstruction, we use the filtered back-projection (FBP) operator, and for MRI, we use the zero-filled inverse Fourier transform. Further, we experiment with two different invertible neural network architectures found in literature: the multi-scale architecture popularized in the \textit{Real NVP} framework \citep{dinh2017density} and an invertible UNet as proposed by Etmann et.\ al.\ \citep{etmann2020iUnets}. Additionally, we propose to use a different base distribution, a radial Gaussian distribution, instead of the widely used standard normal distribution.

\section{Materials and Methods}
In this Section, we introduce normalizing flows and discuss how flow-based models can be implemented. We describe building blocks for invertible neural networks and how they can be used for conditional normalizing flows. In the last part of this Section, we explain the different architectures used for the experiments.

\subsection{Deep Generative Models}
The aim of generative modeling is to build a model using a dataset that represents the underlying distribution of the data. There are two distinct goals in generative modeling. The first is to approximate the probability density function of given samples (i.e.\ density estimation). The second goal is to generate new data samples distributed according to the distribution (i.e.\ sampling). The term deep generative modeling is used when the underlying model is implemented using neural networks. In recent years, a wide variety of powerful methods have been proposed. These can be broadly grouped into latent-variable models, autoregressive models \citep{oord2016pixel, oord2016conditional}, and normalizing flows (NFs) \citep{papamakarios2021normalizing}. The latent-variable models include implicit models, such as generative adversarial networks (GANs) \citep{goodfellow2014generative} and variational autoencoders (VAEs) \citep{rezende2014stochastic, kingma2014autoencoding}. These latent-variable models work by specifying a lower-dimensional latent space and learning a conditional distribution to sample from the image space. GANs are trained using a critic or discriminator network in an adversarial scheme. It was recently shown that GANs have the ability to produce realistic-looking images \citep{brock2018large}. However, it is not possible to compute the likelihood with a GAN. VAEs induce a noisy observation model and utilize a lower bound to the exact likelihood function for training. So it is only possible to evaluate an approximation to the exact likelihood. Additionally, the noisy observation model often leads to blurry-looking images. For autoregressive models (ARMs), the joint distribution is factorized into a product of conditional distributions using the product rule. Using this factorization, neural networks are used to model the dependencies. In this way, the likelihood of an ARM can be computed exactly, but sampling from such a model can be slow. Recently score-based generative models were proposed \citep{song2020score}, which are trained to approximate the gradient of the density and rely on Langevin dynamics for sampling. Models based on the concept of NFs have the advantage of allowing exact likelihood calculation, thus offering the possibility to use a maximum likelihood training and a fast sampling procedure. In distinction to VAEs, they are invertible by design and have no reconstruction loss. 

\subsection{Application of Generative Models to Inverse Problems}
Inverse problems can be studied from a statistical point of view \cite{kaipio2005statistical}. In this interpretation, we are interested in the conditional distribution $p(x|y^\delta)$ of the unknown image $x$ given the measurement data $y^\delta$, the so-called posterior. Using Bayes' theorem, this posterior can be decomposed into a prior $p(x)$ and the likelihood $p(y^\delta|x)$:
\begin{align}
    p(x|y^\delta) \propto p(y^\delta|x) p(x)
\end{align}
For a given noise model, the likelihood $p(y^\delta|x)$ can be evaluated using the forward model $\mathcal{A}: X \rightarrow Y$ \citep{Dashti2017}. The prior $p(x)$ encodes information about the image. Deep generative models are usually incorporated in two ways: learning a model for the prior $p(x)$ \citep{asim2020invertible} or learning a model for the full posterior distribution $p(x|y^\delta)$ \cite{ardizzone2021conditional, Denker2020}. To explore the posterior distribution, other point estimates can be used. Commonly, the maximum a posterior (MAP) estimate 
\begin{equation}
  \begin{aligned}
    \hat{x} &= \argmax_{x \in X} p(x|y^\delta) \\
        &= \argmax_{x \in X} \log(p(y^\delta|x)) + \log(p(x))
\end{aligned}  
\end{equation}
or the conditional mean $\mathbb{E}[x|y^\delta]$ is used as reconstruction, and the conditional variance $\text{Var}[x|y^\delta]$ is used to assess the uncertainty. As computing the conditional mean and the conditional variance would require solving a high-dimensional integral, we use an approximation to estimate both moments as
\begin{align}
    \widehat{\mathbb{E}[x|y^\delta]} = \frac{1}{N} \sum_{i=1}^n x_i \quad \text{and} \quad \widehat{\text{Var}[x|y^\delta]} = \frac{1}{n} \sum_{i=1}^N (x_i - \widehat{\mathbb{E}[x|y^\delta]})^2,
\end{align}
with $N$ i.i.d.\ samples $\{ x_i \}$ drawn from the trained model. In our experiments we focus on directly learning a model for the full posterior $p(x|y^\delta)$.

\subsection{Normalizing Flows}
The concept of NFs is based on the work of Tabak and Turner \citep{tabak2013family}. Flow-based models are constructed using two components: a base distribution and an invertible transformation. Let $\mathbf{z}$ be a random variable with a known probability density function $p_\mathbf{z}$. This distribution is called the base distribution and should be simple to evaluate and sample from. The second component is a transformation $T_\theta:X =\R^n \rightarrow Y=\R^n$, which is parametrized by $\theta$. This transformation has to be invertible, and both $T_\theta$ and $T_\theta^{-1}$ have to be differentiable. This particular class of functions is called a diffeomorphism. The base distribution $p_\mathbf{z}$ induces a distribution, via the invertible transformation $T_\theta$, on the image space $\mathbf{x} = T_\theta(\mathbf{z})$. Using the change-of-variable theorem, it is possible to evaluate the likelihood of this induced distribution:
\begin{align}
\label{eq:changeofvar1}
    p_\theta(x) = p_\mathbf{z}(T_\theta^{-1}(x)) |\det J_{T_\theta^{-1}}(x)|.
\end{align}
Here, $J_{T_\theta^{-1}}(x)$ denotes the Jacobian of $T_\theta^{-1}$. In some cases, it may be of advantage to express \eqref{eq:changeofvar1} using the Jacobian of $T_\theta$: 
\begin{align}
    \label{eq:changeofvar2}
    p_\theta(x) = p_\mathbf{z}(T_\theta^{-1}(x)) |\det J_{T_\theta}(T_\theta^{-1}(x))|^{-1}.
\end{align}
This exact formulation of the probability density offers the possibility to fit the parameters $\theta$ of the NF using maximum likelihood estimation \citep{dinh2015nice}. Assume that we have a dataset of i.i.d.\ samples $\{ x^{(i)} \}_{i=1}^N$ from an unknown target distribution, then this objective is used for training the NF:
\begin{equation}
\begin{aligned}
    \max \L(\theta) &= \sum_{i=1}^N \log(p_\theta(x^{(i)}))  \\
    &= \sum_{i=1}^N \left( \log p(T_\theta^{-1}(x^{(i)})) +  \log |\det J_{T_\theta}(T_\theta^{-1}(x^{(i)}))|\right).
\end{aligned}
\end{equation}
This maximum likelihood objective is equivalent to minimizing the Kullback-Leibler divergence between the unknown target distribution and the induced distribution of the flow-based model \citep{papamakarios2021normalizing}.

The key challenge is to build an expressive invertible transformation $T_\theta$. For this purpose, two essential properties of diffeomorphisms can be exploited. Diffeomorphisms are composable, i.e.\ if $T_1$ and $T_2$ are invertible and differentiable, then the same holds for $T_2 \circ T_1$. Further, it is possible to decompose the computation of the inverse and the Jacobian determinant:
\begin{align}
    (T_2 \circ T_1)^{-1} = T_1^{-1} \circ T_2^{-2} \text{  and  } \det J_{T_2 \circ T_1}(z) = \det J_{T_2}(T_1(z)) \cdot \det J_{T_1}(z)
\end{align}
This allows us to build a complex transformation as a concatenation of simple transformations. We start by defining a base distribution for $\mathbf{z_0}$. Using the concatenated $T_\theta = T_K \circ \dots \circ T_1$, we can compute the probability density of $\mathbf{x} = \mathbf{z_K} = T_\theta(\mathbf{z_0})$ via 
\begin{align}
    p_\theta(z_K) = p_{\mathbf{z_{0}}}(T_\theta^{-1}(z_K)) \prod_{k=1}^K | \det J_{T_k}(T_k^{-1}(z_k)) |^{-1} 
\end{align}
with $z_{k-1} = T_k^{-1}(z_k)$. This composition of transformations leads to the name normalizing flow \citep{papamakarios2021normalizing}. The transformations $T_i$ are a critical part of this formulation. We need transformations that 
\begin{itemize}
    \item are easily invertible, 
    \item offer an efficient calculation of the logarithm of the Jacobian determinant,
\end{itemize}
and are still expressive enough to approximate complex distributions.  Several different models  offer invertibility and tractable determinants, e.g.\ planar flows \citep{rezende2016variational}, residual flows \citep{behrmann2019invertible, chen2020residual}, or Sylvester flows \citep{berg2018sylvester}. We focus on a class of models that are based on so-called coupling layers \citep{gomez2017reversible, dinh2015nice}. 

\subsection{Invertible Neural Networks}
Invertible neural networks consist of layers that guarantee an invertible relationship between their input and output. Therefore, they are ideally suited to be used as normalizing flow. There is also the advantage that the intermediate activations do not have to be stored during backpropagation in training. Compared to regular neural networks, the memory consumption decreases considerably, so more extensive networks or batch sizes can be realized. For both CT \cite{etmann2020iUnets} and MRI \cite{putzky2019invert}, there are already invertible architectures that actively use this property. 

The main building blocks of invertible neural networks used in this work are the so-called coupling layers \citep{gomez2017reversible, dinh2015nice}. Coupling layers are invertible by design and have block triangular Jacobians, which allow for an efficient calculation of the logarithm determinant. The main idea of a coupling layer is that the input is split into two parts, where one part is transformed, whereas the other is left unchanged. It is crucial to implement some mixing or permutation between coupling layers for all dimensions to influence one another. In imaging applications, invertible spatial downsampling operations are also integrated into the network \cite{dinh2017density, kingma2018glow, Ardizzone2019b, etmann2020iUnets}. 

\subsubsection{Coupling Layers}
 Let $x \in \R^n$ and $I_1$, $I_2$ disjoint partitions of $\{1, \dots, n \}$ with $|I_1| = d$ and $|I_2| = n - d$. Then a coupling layer is defined via 
\begin{align}
\begin{array}{rl}
    y_{I_1} &= x_{I_1} \\
    y_{I_2} &= G(x_{I_2}, M(x_{I_1})),
\end{array}
\end{align}
where $G: \R^{n-d} \times \R^{n-d} \rightarrow \R^{n-d}$ is called the coupling law, which has to be invertible w.r.t.\ the first argument. The function $M:\R^d \rightarrow \R^{n-d}$ is the coupling function, which does not need to be invertible and can be implemented as an arbitrary neural network. Two main types of coupling functions have been studied in the literature: additive coupling functions and affine coupling functions. Additive coupling, as used in \citep{dinh2015nice}, follow this design: 
\begin{align}
\label{eq:additiveCoupling}
	\begin{array}{rl}
	y_{I_1} &= x_{I_1} \\
	y_{I_2} &= x_{I_2} + M(x_{I_1}) \\
	\end{array} \Leftrightarrow \begin{array}{rl}
	x_{I_1} &= y_{I_1} \\
	x_{I_2} &= y_{I_2} - M(y_{I_1}). \\
	\end{array}
\end{align}
A more flexible type of coupling is affine coupling \citep{dinh2017density}. Affine coupling layers introduce an additional scaling function to the translation of the additive coupling layer. In this way, a scale $s(x)$ and a translation $t(x)$ are learned, i.e.\ $M(x) = [s(x), t(x)]$:
\begin{align}
\label{eq:affineCoupling}
	\begin{array}{rl}
	y_{I_1} &= x_{I_1} \\
	y_{I_2} &= x_{I_2} \odot \exp(s(x_{I_1})) + t(x_{I_1}) \\
	\end{array} \Leftrightarrow \begin{array}{rl}
	x_{I_1} &= y_{I_1} \\
	x_{I_2} &= \exp(-s(y_{I_1})) \odot (y_{I_2} - t(y_{I_1})) \\
	\end{array}
\end{align}

Instead of choosing $\exp(\cdot)$, sometimes other functions which are non-zero everywhere are used. Because one part of the input is unchanged during the forward pass of a coupling layer, we get a lower block triangular structure for the Jacobian matrix:

\begin{align}
    \frac{\partial y}{\partial x} = \begin{pmatrix}
        I_m &  0 \\
    \frac{\partial y_{I_2}}{\partial x_{I_1}}     & \frac{\partial y_{I_2}}{\partial x_{I_2}}
    \end{pmatrix}.
\end{align}

This allows us to compute the determinant as $\det\left(\frac{\partial y}{\partial x}\right) = \det\left(\frac{\partial y_{I_2}}{\partial x_{I_2}}\right)$, which drastically reduces the computational complexity. For additive coupling layers, this further reduces to the identity matrix, i.e.\ they have a unit determinant. Affine coupling layers have a diagonal structure in the block: 
\begin{align}
    \det\left(\frac{\partial y_{I_2}}{\partial x_{I_2}}\right) = \exp\left(\sum_{i \in I_1} s(x_1)_i)\right).
\end{align}
However, as $s(x_1)$ is already evaluated in the forward pass, computing the determinant does not involve significant computational effort. The special structure of the Jacobian highlights the fact that some parts of the input are not processed and have no influence on each other. It is essential to include some permutation or mixing of dimensions in order to build an expressive sequence of coupling layers.

\subsubsection{Channel Mixing and Downsampling}
\begin{figure}
    \centering
    \includegraphics[scale=0.15]{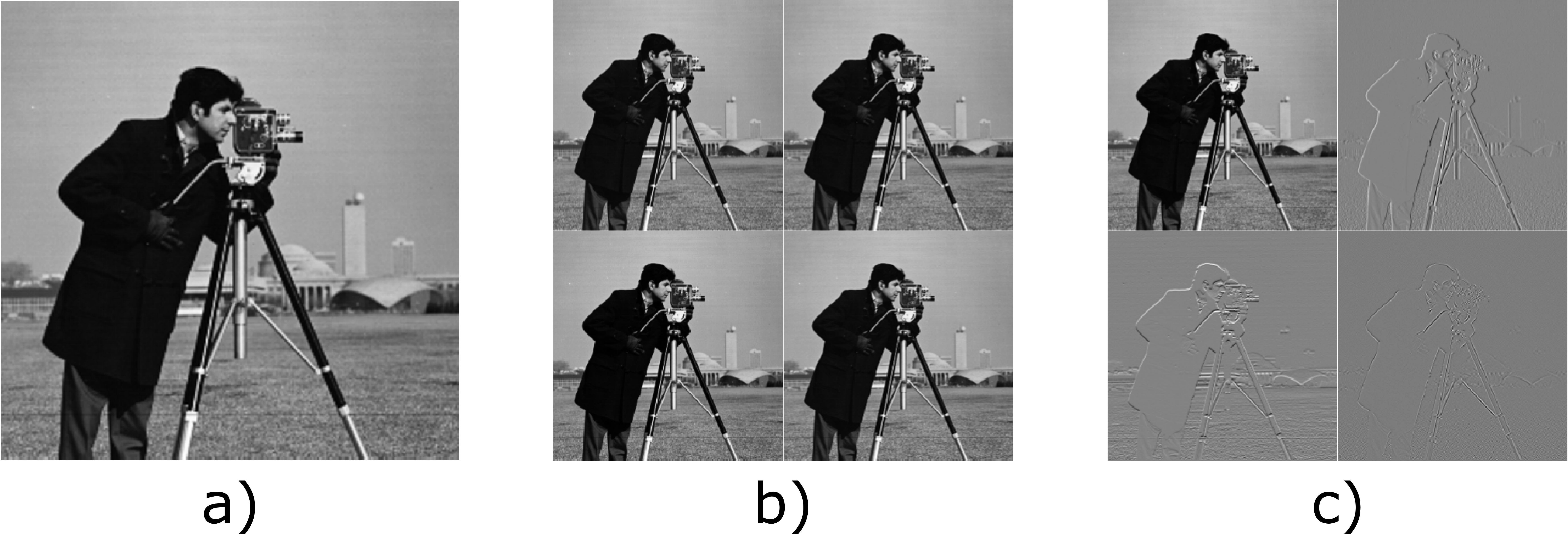}
    \caption{Input image $a)$ and output of checkerboard downsampling $b)$ and haar downsampling $c)$. Adapted from \citep{etmann2020iUnets}.}
    \label{fig:haarvsshuffle}
\end{figure}

For each coupling layer, the input is split into two parts, and only one-half is processed. For image data, this splitting usually is done in the channel dimension. Let $u \in \R^{c \times h \times w}$ be an image with $c$ channels. We choose $c_1, c_2$ such that $c_1 + c_2 = c$. The image is then split into two parts $u_{I_1} \in \R^{c_1 \times h \times w}$ and $u_{I_2} \in \R^{c_2 \times h \times w}$. In earlier works, the permutation after each coupling layer was implemented as a fixed random channel shuffling \citep{dinh2015nice}. In the \textit{Glow} architecture, an improvement was seen when using fixed $1 \times 1$ convolutions instead of simple permutations \cite{kingma2018glow}. These fixed convolutions can be seen as a generalization of random shuffling. Another central part of invertible neural networks in imaging applications is invertible downsampling operations, i.e.\ reduction of the spatial dimensions of image data. The standard downsampling operations in CNNs, like pooling layers or strided convolutions, are inherently non-invertible as they reduce the dimensionality of the image. Invertible downsampling operations reduce the spatial dimension while simultaneously increasing the number of channels, thus keeping the overall dimensionality the same. Let $u \in \R^{c \times h \times w}$ be an image with $c$ channels, where both the height $h$ and the width $w$ are even. An invertible downsampling operation halves both spatial dimensions and quadruples the number of channels, i.e.\ $\Tilde{u} \in \R^{4c \times h/2 \times w/2}$. There are three main types of invertible downsampling operations used in the literature. The first is checkerboard downsampling, which is a simple rearrangement of the image pixels \cite{jacobsen2018irevnet}. A more advanced type of downsampling is haar downsampling introduced in \cite{Ardizzone2019b}, which uses the 2D haar transform to decompose the image into average channels and vertical, diagonal, and horizontal components. These two downsampling operations are illustrated in Figure \ref{fig:haarvsshuffle}. Recently Etmann et al.\ introduced a learnable invertible downsampling operation \citep{etmann2020iUnets}. 

\subsection{Base Distribution}
\label{sec:base_distribution}
In most applications, a standard $n$-dimensional Gaussian $\mathbf{z} \sim \mathcal{N}(0, I)$ is chosen as the base distribution, which leads to the following log-likelihood: 
\begin{align}
    \log(p_\mathbf{z}(z)) = - \frac{1}{2} \| z \|_2^2 - \frac{n}{2} \log(2\pi).
\end{align}
The second term is constant w.r.t.\ $z$ and can be dropped during training. It has been observed that the likelihood of flow-based models sometimes exhibits artifacts, i.e.\ out-of-distribution data is often assigned a higher likelihood than training data \citep{nalisnick2018deep}. In \citep{nalisnick2019detecting}, the authors suggest that this behavior is due to the difference between the \textit{high likelihood set} and the \textit{typical set} in high-dimensional Gaussian distributions. For a standard Gaussian, the region of highest density is at its mean, but the typical set is at a distance of $\sqrt{d}$ away from the mean. In \citep{farquhar2020radial}, the authors address this problem for Bayesian Neural Networks and choose a \textit{radial} Gaussian distribution where the typical set and high-density region coincide. This radial Gaussian is formulated in hyperspherical coordinates, where the radius is distributed according to a half-normal distribution, i.e.\ $r = \abs{\hat{r}}$ with $\hat{r} \sim \mathcal{N}(0,1)$, and all angular coordinates follow a uniform distribution over the hypersphere. We use this radial distribution as a base distribution for training flow-based models. This radial distribution leads to the following log-likelihood
\begin{align}
        \ln p_\mathbf{z}(z) = \ln\left(\frac{\sqrt{2}}{\sqrt{\pi} S_n}\right) - (n-1) \ln(\| z \|_2) - \frac{\| z \|_2^2}{2},
\end{align}
where $S_n$ is the surface of the $n$ dimensional unit sphere. The derivation can be found in Appendix~\ref{app:radial_distr}. Sampling is nearly as efficient as for the standard Gaussian distribution. First, a point $x \sim \mathcal{N}(0, I_n)$ is sampled and normalized. This point is then scaled using a radius $r = |\hat{r}|$ with $\hat{r} \sim \mathcal{N}(0,1)$.

\subsection{Conditional Normalizing Flow}
Let $\mathbf{x}$ and $\mathbf{y}$ be two random variables over two spaces $X$ and $Y$. For our applications, we always use $X = \R^n$ and $Y = \R^m$. The goal of conditional density estimation is to approximate the conditional probability distribution $p(x|y)$ given an i.i.d.\ data set $\{ (x^{(i)}, y^{(i)}) \}_{i=1}^N$ of input-output pairs sampled from the joint distribution $p(x,y)$. We use a conditional normalizing flow (CNF) to build a probabilistic model $p_\theta(x|y)$ to approximate the unknown conditional distribution $p(x|y)$ \citep{winkler2019learning, Ardizzone2019b}. A CNF consists of a transformation $T_\theta: Z \times Y \rightarrow X$ that has to be invertible w.r.t.\ the first argument and both $T_\theta(\cdot; y)$ and $T_\theta^{-1}(\cdot; y)$ have to be differentiable for every $y \in Y$. By choosing a base distribution $p_\mathbf{z}$, the CNF model induces a probability distribution, and the density can be evaluated via the change-of-variable method: 
\begin{align}
    p_\theta(x|y) = p_\mathbf{z}(T_\theta^{-1}(x;y)) \left| \det\left( \frac{\partial T_\theta^{-1}(x;y)}{\partial x}\right)\right|.
\end{align}
We use $J_{T_\theta^{-1}}(x;y) = \frac{\partial T_\theta^{-1}(x;y)}{\partial x}$ as a shorthand notation for the Jacobian matrix. Fitting the parameters $\theta$ of the CNF can be done using a maximum likelihood loss:
\begin{equation}
\begin{aligned}        
\max_\theta \L(\theta) &= \sum_{i=1}^N \log(p_\theta(x^{(i)} | y^{(i)}))  \\
    &= \sum_{i=1}^N \left( \log p(T_\theta^{-1}(x^{(i)}; y^{(i)})) +  \log\left(|\det J_{T_\theta^{-1}}(x^{(i)}; y^{(i)})|\right) \right).
\end{aligned}
\end{equation}
We use the same trick as for the NF and implement the CNF as a concatenation of simple invertible building blocks. 

\subsubsection{Conditional Coupling Layers}
Conditional coupling layers are the primary way of constructing expressive CNF models. They can be seen as an extension of the original coupling layers and were introduced \citep{Ardizzone2019a} for modeling conditional image densities. For a conditional coupling layer, we extend the coupling function $M$ to take the measurements $y^\delta$ as an additional input. Let $x \in \R^n$ be the input, $y^\delta \in R^m$ the measurements, and $I_1$, $I_2$ disjoint partitions of $\{1, \dots, n \}$ with $|I_1| = d$ and $|I_2| = n - d$. Then a conditional coupling layer is defined by 
\begin{align}
\begin{array}{rl}
    y_{I_1} &= x_{I_1} \\
    y_{I_2} &= G(x_{I_2}, M(x_{I_1}, y^\delta))
\end{array}
\end{align}
where $G: \R^{n-d} \times \R^{n-d} \rightarrow \R^{n-d}$ is called the coupling law, which has to invertible w.r.t.\ the first argument. Function $M:\R^d \times \R^m \rightarrow \R^{n-d}$ is the coupling function. Conditional coupling layers offer the same advantages as regular coupling layers, i.e.\ a block triangular Jacobian and analytical invertibility. In our experiments, we use mainly conditional affine coupling layer, i.e.\ replacing $s(x_{I_1})$ and $t(x_{I_1})$ with $s(x_{I_1}, y^\delta)$ and $t(x_{I_1}, y^\delta)$. For any fixed conditional input $y^\delta$, the conditional normalizing flow is invertible.

Another way of introducing the conditional input $y^\delta$ into the model is to use a conditional base distribution \citep{winkler2019learning}. In this approach, the base distribution can be modeled as a normal distribution where the mean and variance are functions of $y^\delta$, i.e.\ $p(\mathbf{z}|y^\delta) = \mathcal{N}(\mathbf{z}; \mu(y^\delta), \sigma^2(y^\delta))$. Both the mean and variance function can be parametrized as a neural network and trained in parallel to the flow-based model.

\subsection{Conditioning Network}
\label{subsec:conditioning}
Instead of directly using the measurements $y^\delta$ as an additional input to the conditional coupling layer, a conditioning network $H$ is used, which transforms the $y^\delta$ to $h = H(y^\delta)$ \citep{Ardizzone2019a, winkler2019learning}. The motivation behind this is that the conditioning network can learn to extract essential features. This decouples the feature extraction and the density modeling. It is possible to either use a fixed, pre-trained network $H$ or train the conditioning network parallel to the CNF. This conditioning network is often implemented as a big CNN. As convolutional networks are built to exploit equivariance in natural images, they are not ideally suited for CT or MRI measurement data. Instead, we implemented this conditioning network as a model-based inversion layer $\mathcal{A}^\dagger$, which maps from the measurement space to the image space, concatenated with a post-processing CNN to extract features from this initial reconstruction.

Depending on the structure of the conditioning network, an additional loss term for this network can be used during training. One option is to compare the output of $H$ to the ground truth data and thereby train a second reconstruction path within the whole cINN. The goal is to get a single high-quality reconstruction from the conditioning network and cover the uncertainties, e.g.\ from ambiguous solutions, in the sampled reconstruction from the CNF. During inference, the output from the conditioning network and the CNF can be combined to create the final reconstruction.

\subsection{Multi-scale Architecture}
\begin{figure}
    \centering
    \includegraphics{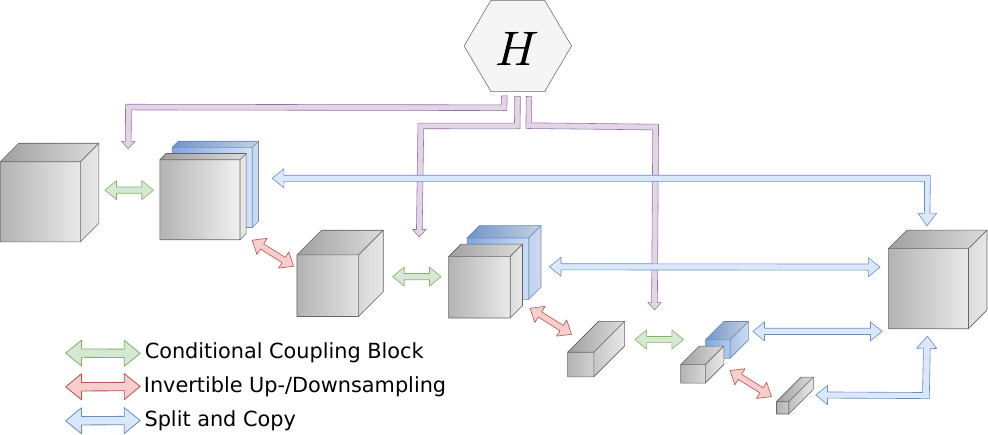}
    \caption{Multi-scale Architecture with conditioning network $H$. The conditioning network processes the conditioning input $y^\delta$ and outputs this to the respective conditional coupling layer.}
    \label{fig:Multiscale}
\end{figure}

Unlike other latent variable models, such as GANs or VAEs, flow-based models work with a full-dimensional base distribution. This is necessary to ensure bijectivity. However, it is also expensive, both in memory cost and in computational complexity, to propagate the full-dimensional image through the network. A typical architecture for flow-based models is the \textit{multi-scale architecture} \citep{dinh2017density}. This architecture combines coupling blocks, downsampling, and splitting operations. A part of the intermediate representation is split off and directly forwarded to the output for each scale. This combination of splitting and feed-forwarding creates a hierarchy of features and reduces the computational effort. We visualize this architecture in Figure \ref{fig:Multiscale}. In our experiments, we always use downsampling of factor \num{2} after each scale. A multi-scale architecture with $L$ scales can be described by:
\begin{align*}
    x^{0} &= x \\
    (z^{i+1}, x^{i+1}) &= f^{i+1}(x^{i}, H^{i}(y^\delta)) \\
    z^{L} &= f^{L}(x^{L-1}, H^{L-1}(y^\delta)) \\ 
    z &= (z^1, \dots z^L).
\end{align*}
Each $f^{i}$ consists of a coupling $\rightarrow$ downsampling $\rightarrow$ coupling $\rightarrow$ splitting operation.

The \textit{multi-scale architecture} follows the NICE and Real-NVP framework \cite{dinh2015nice, dinh2017density} and is related to the i-RevNet architecture \citep{jacobsen2018irevnet}. However, in i-RevNet, the authors refrain from splitting the dimensions in their bijective architecture.

\subsection{Invertible UNet}
\label{sec:iUnet}
\begin{figure}
    \centering
    \includegraphics{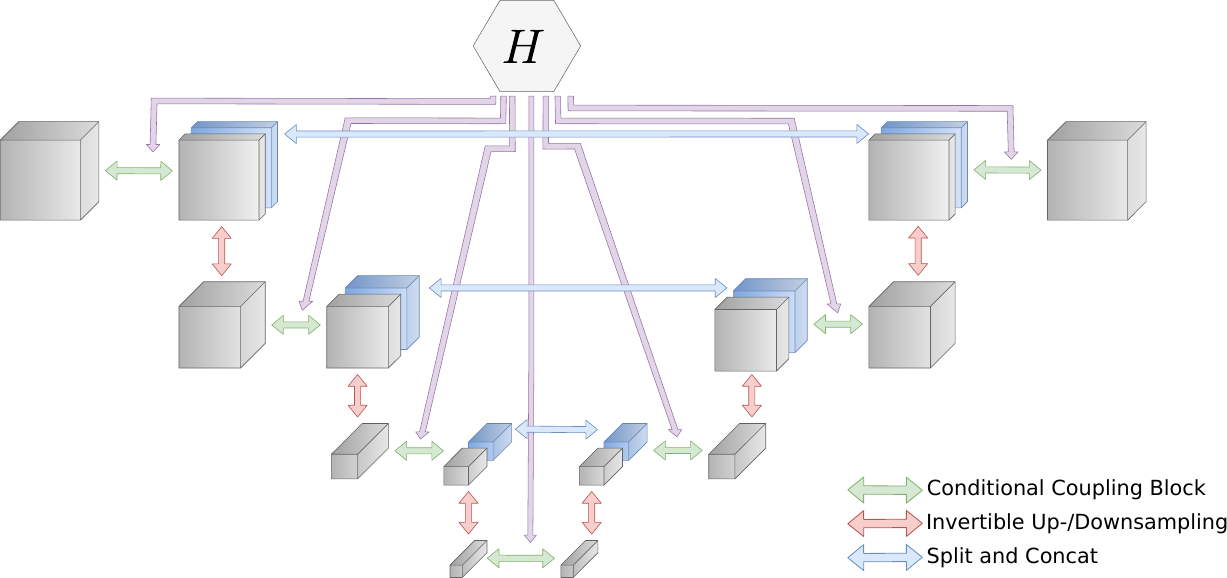}
    \caption{End-to-end invertible UNet with conditioning network $H$. The conditioning network processes the conditioning input $y^\delta$ and outputs this to the respective conditional coupling layer.}
    \label{fig:iUNet}
\end{figure}

 With the iUNet, we follow the work of Etmann et al.\ \cite{etmann2020iUnets}. The idea is to adapt the concept of the UNet architecture \cite{ronneberger2015unet} and replace all common layers with their invertible counterparts. In addition, we introduce a conditioning network $H$, which also has a UNet structure. In this case, the layers do not have to be invertible. Network $H$ uses the same spatial down- and upsampling scales as the iUNet. At each scale, the current activation $H^i_{u;d}$ is used as conditioning for the respective block $f^{i+1}_{d;u}$ in the iUNet. Note that the direction of the UNet is inverse to the iUNet since it starts from measurement $y^\delta \in Y$ and maps to $X$. A representation of the whole network is shown in Figure \ref{fig:iUNet}. For an architecture with $L$ scales, we get:
\begin{align*}
    x_d^{0} &= x \\
    (c^{i+1}, x_d^{i+1}) &= f_d^{i+1}(x_d^{i},    H_u^{i}(y^\delta)),\quad &&i = 0, \dots, L-2 \\
     x_d^{L} &= f_d^{L}(x_d^{L-1}, H_u^{L-1}(y^\delta)) \\ 
     x_u^{L} &= x_d^{L} \\
     x_u^{i-1} &= f_u^{i}((x_u^{i}, c^i),  H_d^{i}(y^\delta)),\quad &&i = L, \dots, 1 \\
     z &= x_u^{0}
\end{align*}    
where indices $d,u$ denote the down- and upsampling path, respectively. Block $f_d^i$ consists of coupling $\rightarrow$ downsampling $\rightarrow$ split, $f_d^{L}$ is just coupling and
$f_u^i$ is upsampling $\rightarrow$ concat $\rightarrow$ coupling. Compared with the multi-scale architecture, the iUNet concatenates the splits step-by-step in the upsampling path and not all together in the last layer.

The conditioning UNet $H$ creates outputs in the image domain $X$. Therefore, we can introduce an additional conditioning loss, as proposed in Section~\ref{subsec:conditioning}. Specifically, we use
\begin{align}
\label{eq:cond_loss_mse}
    \min_\theta -\log(p_\theta(x \vert y^\delta)) + \alpha \operatorname{MSE}(H(y^\delta), x),
\end{align}
where $\alpha \geq 0$ is a weighting factor. Note that one can also use a pre-trained UNet with fixed parameters as conditioning and benefit from the advantages of the CNF in comparison to a simple post-processing approach.

\section{Experimental Setup}
In this section, we present three different applications used to evaluate different architectures for conditional flow-based models. In the first example, we study compressed sensing with Gaussian measurements on the popular MNIST dataset \citep{lecunmnist}. The other two applications cover essential aspects of medical imaging: accelerated magnetic resonance imaging and low-dose computed tomography. In these two medical imaging scenarios, different sources introduce uncertainty to the reconstruction process. We have an undersampling case in accelerated MRI, w, i.e.\ we have fewer measurements than necessary according to the  Nyquist-Sampling theorem. So, a strong prior is needed for a good reconstruction. The challenge in low-dose CT is that the lower radiation dose leads to a worse signal-to-noise ratio. Although we are in an oversampling case, the reconstruction is complicated by a more significant amount of noise.

Our source code is publicly available at \url{https://github.com/jleuschn/cinn_for_imaging}.

\subsection{Compressed Sensing}
\label{sec:compressed_sensing}
As an initial example, we study a similar setup as in \cite{genzel2020solving}. The goal is the recovery of an image from Gaussian measurements. We evaluate our models on the popular MNIST \citep{lecunmnist} dataset, which consists of $28 \times 28$-size images of handwritten digits. MNIST contains \num{60000} training images and \num{10000} test images. We split the \num{60000} training images into \num{50000} for training the CNF model and \num{10000} for validation. The forward operator is matrix $\mathcal{A} \in \R^{m \times n}$. It has independent Gaussian entries with zero mean and variance $1/m$, i.e.\ $\mathcal{A}_{i,j} \sim \mathcal{N}(0, 1/m)$. We use $m=196, n=784$, i.e.\ \num{4} times downsampling. We added $10\%$ relative noise to the simulated measurements. In this experiment, we want to study the influence of the inversion layer in the conditioning network $H$. We use the generalized inverse $\mathcal{A}^\dagger = \mathcal{A}^+$ and a TV-regularized solution $\mathcal{A}^\dagger = (\mathcal{A}^T \mathcal{A} + \lambda \nabla^T \nabla) \mathcal{A}^T$ with a regularization parameter $\lambda=0.02$. We further use the same neural network architecture for both the conditional invertible network and the conditioning network for both choices of $\mathcal{A}^\dagger$. The cINN was implemented as a multi-scale architecture with two learnable downsampling operations. The exact implementation can be found in Appendix \ref{app:comp_sense_arch}.

\subsection{Computed Tomography}
When describing the propagation of radiation through biological tissue, two processes have to be considered: absorption and scattering. For high-energy X-ray beams, the scattering effect is usually neglected. The forward problem in parallel-beam computed tomography can then be described by the 2D Radon transform \citep{radon1986determination}: 
\begin{align}
    Ax(s,\varphi) = \int_\R x\left( s \begin{bmatrix} \cos(\varphi) \\ - \sin(\varphi) \end{bmatrix} + t 
    \begin{bmatrix} -\sin(\varphi) \\ \cos(\varphi) \end{bmatrix}\right) \dd t,
\end{align}
where $x$ is the spatial varying mass absorption coefficient, which depends on tissue type and density. The Radon transform corresponds to the log-ratio between the source intensity and the measured intensity. 

For continuous, noise-free measurements, the filtered back-projection (FBP) in combination with the Ram-Lak filter gives the exact inversion formula \citep{buzug2008computed_tomography}. In general, recovering the image is a mildly ill-posed problem in the sense of Nashed \citep{nashed1987ill_posed, natterer2001mathematics}. This means that slight deviations in the measurement, e.g.\ noise, can lead to significant changes in the reconstruction. The influence of the noise can be reduced by choosing an adequate filter for the FBP. Another challenge arises from the discretization of real measurements, which can lead to artifacts in the FBP reconstruction. Over the years, a number of different reconstruction methods, like algebraic reconstruction techniques \cite{gordon1970} (ART) and total variation (TV) regularization \cite{sidky2008image}, were introduced to compensate for the drawbacks of the FBP. Recently, deep learning approaches extended the choice of methods to push the boundaries on image quality for low-dose, sparse-angle, and limited-angle measurements \cite{adler2018pd, leuschner2021quantitative, jin2017deep, bubba2021}. 

\begin{figure}
    \centering
    \includegraphics[scale=0.425]{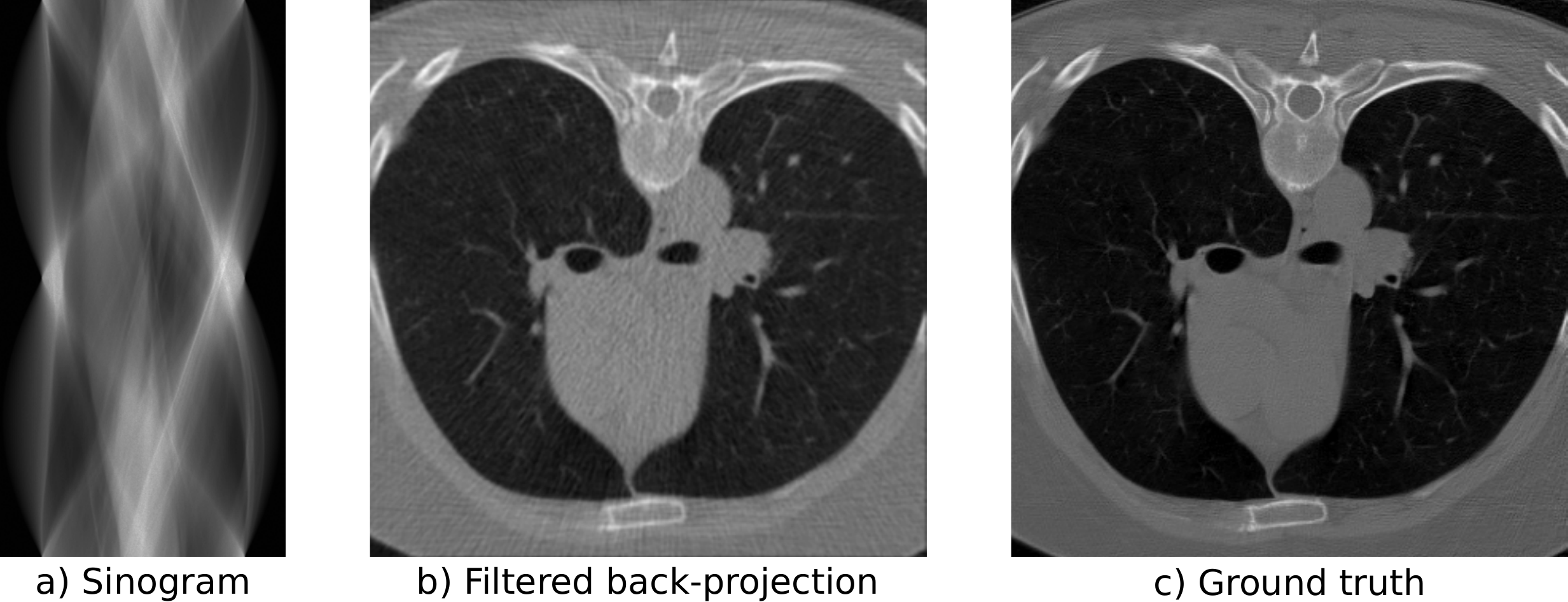}
    \caption{Reconstruction and measurements for the low-dose  \textit{LoDoPaB-CT} data.}
    \label{fig:lodopab_measurements}
\end{figure}

In our experiments, we use the LoDoPaB-CT dataset \cite{Leuschner_2021} to replicate the challenges that arise from low-dose CT measurements. The dataset contains over \num{40000} normal-dose, medical CT images from the human thorax from around \num{800} patients. Poisson noise is used to simulate
the corresponding low-dose measurements. See Figure \ref{fig:lodopab_measurements} for an example of a simulated low-dose measurement, an FBP reconstruction, and the ground truth image.
LoDoPaB-CT has a dedicated test set that we use for the evaluation and comparison of our models. In addition, there is a special challenge set with undisclosed ground truth data. We evaluate the best model from our experiments on this set to allow for a comparison with other reconstruction approaches. The challenge results can be found on the online leaderboard (\url{https://lodopab.grand-challenge.org/evaluation/challenge/leaderboard/}).

\subsection{Magnetic Resonance Imaging}
We will now briefly introduce MRI and the considered simple model, following the description in \cite{doneva2020mathematical}, to which we refer the reader for more details, including limitations of the model.

In MRI, one measures the radio frequency (RF) responses of nuclei (e.g.\ protons) to RF pulses while applying different external magnetic fields in order to obtain a density image.
A strong static magnetic field is applied that causes the resonance frequency of the nuclei to be within the RF range.
Pulses at this frequency are emitted using an RF transmitting coil, triggering RF response signals detected by an RF receiving coil.
For spatial encoding, configurable magnetic gradient fields $G=(G_x,G_y,G_z)$ are applied that change the applied magnetic field and thereby the resonance frequency depending on the location.
During a scan, different gradient fields $G$ are selected for each repetition of a pulse sequence.

A simple model for the measured receive coil signal in each repetition is given by
\begin{align*}
    y(t) = \int x(r) \exp(-2 \pi i k(t) \cdot r) \dd r, \qquad k(t) = \gamma \int_0^t G(\tau) \dd \tau,
\end{align*}
where $x$ is the spatial signal density (i.e.\ the image) and $k$ specifies a position in the so-called $k$-space, which coincides with the Fourier space.
The choice of $G$ determines the trajectory of $k$ for this repetition.
By collecting samples from multiple repetitions, one can obtain a complete Cartesian sampling of the $k$-space that satisfies the Nyquist-Shannon sampling theorem.
This enables (approximate) reconstruction via the inverse fast Fourier transform (IFFT).

A major limiting factor is the time-consuming measurement process, which directly depends on the number of repetitions required to obtain a full sampling of the $k$-space.
While using fewer repetitions accelerates the process, it leads to an underdetermined reconstruction problem and can introduce artifacts due to the missing frequencies.
In order to reconstruct from undersampled measurement data, prior information needs to be incorporated.
Additionally, measurements are noisy in practice, further increasing reconstruction ambiguity since all solutions matching the measured data within the noise level would be plausible.
This strengthens the requirement of prior information.

\begin{figure}
    \centering
    \includegraphics[scale=0.425]{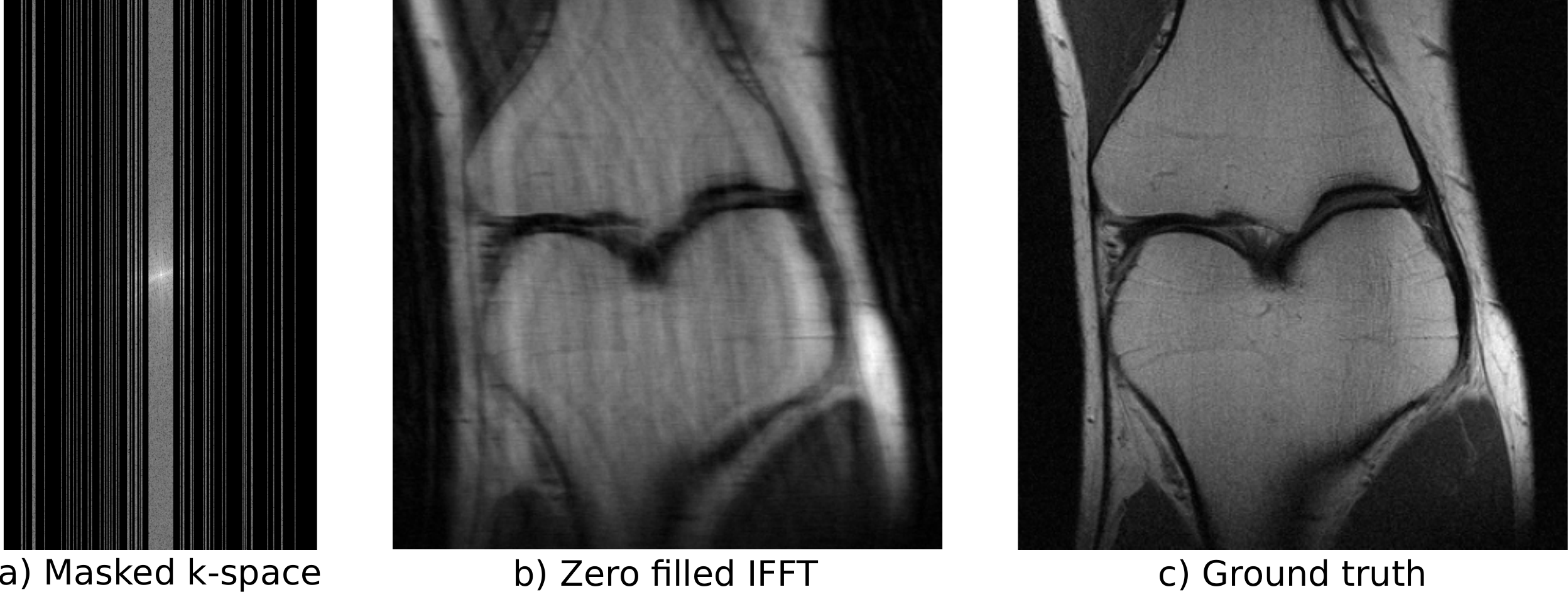}
    \caption{Measurements and reconstruction for the single-coil \textit{fastMRI} data.}
    \label{fig:fastmri_measurements}
\end{figure}

In our experiments, we used the emulated single-coil measurements from the NYU fastMRI database \citep{knollfastmri,zbontar2019fastmri}. The fully sampled measurements were retrospectively subsampled to simulate accelerated MRI data. See Figure \ref{fig:fastmri_measurements} for an example of a subsampled measurement, a zero-filled IFFT reconstruction, and the ground truth obtained from the full measurement. We used an acceleration factor of $4$, i.e.\ only $25 \%$ of frequencies were kept. Undersampling was performed by selecting $8 \%$ of the lowest frequencies and randomly adding higher frequencies until the acceleration factor was reached. The public dataset consists of a training part and a validation part. In total, the training dataset includes \num{973} volumes (\num{34742} slices) and the validation dataset \num{199} volumes (\num{7135} slices). Additionally, there is a private test set that consists of \num{108} volumes (\num{3903} slices). For this private test set, only the undersampled measurements are available, and the models can only be evaluated on the official fastMRI website (\url{https://fastmri.org/}). Our best model can be found on the public leaderboard for ``Single-Coil Knee'', allowing for comparison with other approaches (our submission is named ``cINN v2''). The fastMRI dataset includes scans from two different pulse sequences: coronal proton-density weighting with (PDFS) and without (PD) fat suppression. We trained our models on the full dataset but used the distinction into PD and PDFS for evaluation on the validation set.

\section{Results}
In this section, we present the results of the three different experimental setups. The focus here is on LoDoPaB-CT and fastMRI. For these use cases, we compare different architectures and ablations during training. To assess the performance, we evaluate the peak-signal-to-noise ratio (PSNR) and the structural similarity index measure (SSIM) \citep{wang2004image} on the datasets. The PSNR is strongly related to the mean squared error and expresses the ratio of the maximum possible value to the reconstruction error. In general, a higher PSNR corresponds to a better reconstruction. The SSIM compares the overall image structure, including luminance and contrast, of the reconstruction and the ground truth image. A detailed definition of the evaluation metrics can be found in Appendix~\ref{app:eval_metrics}.

\subsection{Compressed Sensing}
\label{sec:results:comp_sense}
\begin{figure}[h]
    \centering
    \includegraphics[scale=0.4]{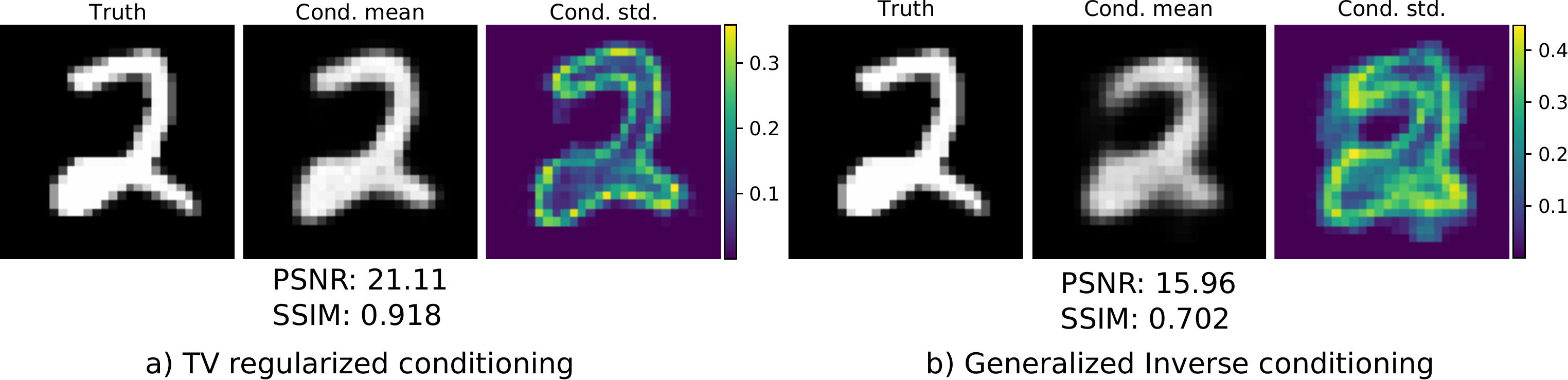}
    \caption{Conditioned mean and standard deviation for the different inversion layers.}
    \label{fig:MNIST_condMean}
\end{figure}

\begin{figure}[h]
    \centering
    \includegraphics[scale=0.5]{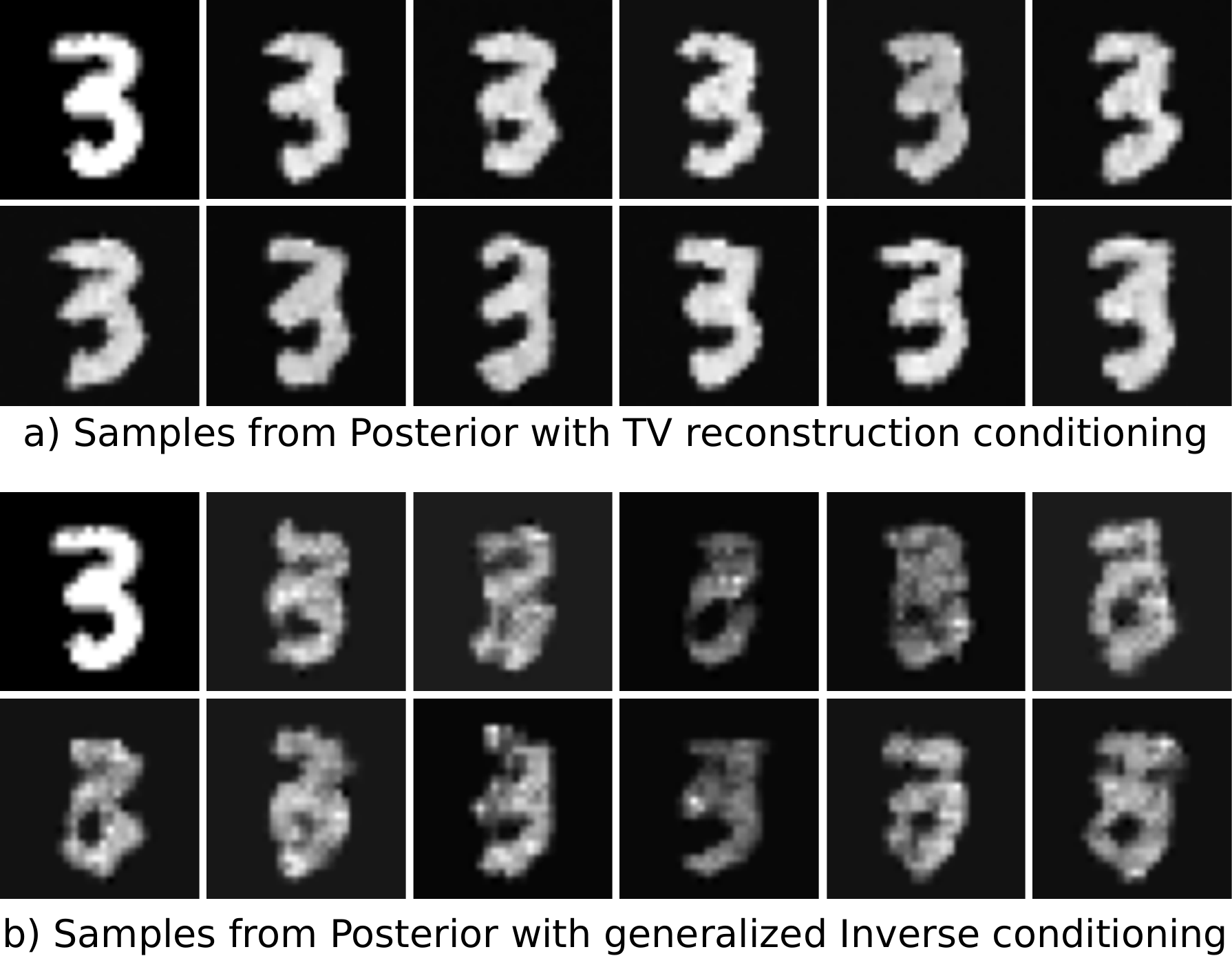}
    \caption{Samples from the Posterior learned by the cINN. The ground-truth sample is shown in the upper left corner. In $a)$, we used the conditioning based on the TV regularized reconstruction, and in $b)$, the conditioning was chosen as the generalized inverse. It can be seen that individual samples from the generalized inverse conditioning do not look realistic.}
    \label{fig:MNIST_samples}
\end{figure}

Both models were trained using the Adam optimizer \cite{kingma2014adam} until convergence with a fixed learning rate of $\num{1e-4}$. The final model was chosen as the best model regarding the negative log-likelihood on the validation set. The conditional mean was used as reconstruction, and we evaluated both the PSNR and SSIM for the entire test set. The results can be seen in Table \ref{tab:results_mnist}. The TV-regularized solution as the conditioning input leads to a drastic improvement both in terms of PSNR and SSIM. A visual comparison of one reconstruction is given in Figure \ref{fig:MNIST_condMean}. One can see that the reconstruction using the TV-regularized solution fits way better to the original ground truth image. Also, the conditioned standard deviation is more centered towards the edges of the number. The reconstruction using the generalized inverse as a conditioning input is much smoother and more blurry. The conditional standard deviation is not so focused on specific features on the image. Lastly, we illustrated samples from both models in Figure \ref{fig:MNIST_samples}. The samples drawn from the model using the TV-regularized conditioning input look much more realistic. 

\begin{specialtable}[h]
\caption{Results for compressed sensing on the MNIST test dataset. Conditioned mean computed with $100$ samples.}
\begin{tabular}{lcc|cc}
\multicolumn{5}{c}{\textit{Compressed Sensing on MNIST}} \\
\toprule        & \multicolumn{2}{c|}{$\mathcal{A}^\dagger = \mathcal{A}^+$} & \multicolumn{2}{c}{$\mathcal{A}^\dagger = (\mathcal{A}^T \mathcal{A} + \lambda \nabla^T \nabla) \mathcal{A}^T$} \\
& PSNR & SSIM  & PSNR & SSIM \\
Multi-scale cINN & $17.32 \pm 2.05$ & $0.752 \pm 0.084$ & \colorbox{gray!20}{$19.89 \pm 2.54$} & \colorbox{gray!20}{$0.868 \pm 0.063$} \\ \bottomrule
\end{tabular}
\label{tab:results_mnist}
\end{specialtable}

\subsection{Computed Tomography}
First, we investigate different conditioning networks for the multi-scale architecture. Based on these results, we compare the multi-scale network to the iUNet. The experiments also include variations in the target distribution and the loss function. The overall results on the LoDoPaB-CT test set are shown in Table \ref{tab:lodopab_results}.

For all comparisons between multi-scale architecture and iUNet, a unified setting was used. Both networks had a similar size ($2.9$ Mio.\ for the iUNet and $3.2$ Mio.\ for the multi-scale architecture).
We used \num{5} scales for all networks. The inversion model inside the conditioning is the filtered back-projection (FBP). For the iUNet additive coupling layers and for the multi-scale architecture, affine coupling layers were used. Gradient descent steps with the Adam optimizer \cite{kingma2014adam}, an  initial learning rate of $\num{1e-4}$, and a reduction factor of \num{0.8} on plateaus were performed during training. The best parameter configuration for each setting was chosen based on the lowest negative log-likelihood on the validation set. 

\subsubsection{Architecture of Conditioning Network}
We tested three different architectures for the conditioning network in the multi-scale cINN model. The first architecture (Average Pooling) consisted of one initial learned convolutional layer to blow up the number of channels followed by average pooling operations to reduce the spatial dimensions to the correct size. In the next architecture (CNN), the one initial convolutional layer was replaced by a fully convolutional neural network. The last architecture (ResNet) used residual connections and replaced all average pooling operations with strided convolutional layers. All models were trained using the same initialization with the Adam optimizer \cite{kingma2014adam}. We evaluated all three choices on the \textit{LoDoPaB} test set, and the results can be seen in Table \ref{tab:lodopab_condnetwork}. In our experiments, increasing the complexity of the conditioning network also increased the reconstruction quality in terms of SSIM and PSNR. We suspect that this increase in quality is related to the fact that a more extensive conditioning network can extract a larger amount and more essential features from the conditioning input. 

\begin{specialtable}[h]
\caption{Influence of the type of conditioning network for the multi-scale cINN. PSNR and SSIM were evaluated on the full \textit{LoDoPaB} test set using $1000$ samples for the cond. mean.}
\begin{tabular}{cccc}
 \multicolumn{4}{c}{\textit{LoDoPaB-CT}} \\
\toprule Model & Cond. Network & PSNR & SSIM  \\
\toprule Multi-scale & Average Pooling & $33.15 \pm 3.64$ & $0.806 \pm 0.156$  \\                                 & CNN & $34.64 \pm 4.18$ & $0.826 \pm 0.160$  \\
                     & ResNet & \colorbox{gray!20}{$35.07 \pm 4.34$} & \colorbox{gray!20}{$0.831 \pm 0.160$}  \\
 \bottomrule \\
\end{tabular}
\label{tab:lodopab_condnetwork}
\end{specialtable}

Based on these results, we chose the ResNet conditioning for the following experiments. Note that we reduced the number of parameters of the multi-scale cINN in the other experiments to be comparable with the iUNet model and shorten the time for training. Overall, this has only a minor effect on the reconstruction quality.

\subsubsection{Base Distribution}
It has been proven that under reasonable conditions for the true density, any base distribution can be used for normalizing flows \citep{papamakarios2021normalizing}. However, the question arises of whether some distributions are more suitable than others. We study two different choices for the base distribution: a standard Gaussian distribution used in most flow-based models and a radial Gaussian as discussed in Section \ref{sec:base_distribution}. As we are interested in the conditional mean in most applications, sample efficiency is vital for the practical implementation and evaluation of a flow-based model. 

\begin{specialtable}
\caption{Results for the \textit{LoDoPaB-CT} test set. Conditioned mean computed with $1000$ samples. Unless stated otherwise, training noise was applied and no cond. loss was used.}
\begin{tabular}{ccccc}
\multicolumn{5}{c}{\textit{LoDoPaB-CT}} \\
\toprule Model & Base Distribution & Train Noise & PSNR & SSIM  \\
\toprule Multi-scale 
& Normal & Yes & \colorbox{gray!20}{$34.99 \pm 4.26$} & \colorbox{gray!20}{$0.830 \pm 0.158$} \\
&        & No  & $34.97 \pm 4.28$ & \colorbox{gray!20}{$0.830 \pm 0.157$} \\
\cmidrule{2-5}       
& Radial & Yes & $34.89 \pm 4.29$ & $0.823 \pm 0.161$  \\
&        & No  & $34.65 \pm 4.25$ & $0.829 \pm 0.161$  \\
\cmidrule{1-5}   iUNet 
& Normal & Yes & $34.69 \pm 4.13$ & $0.806 \pm 0.151$ \\
&        & No  & $34.98 \pm 4.19$ & $0.823 \pm 0.148$ \\
\cmidrule{2-5}       
& Radial & Yes & $34.75 \pm 4.23$ & $0.819 \pm 0.153$ \\
&        & No  & $34.57 \pm 4.40$ & \colorbox{gray!20}{$0.830 \pm 0.158$} \\
\bottomrule  \toprule  & & Cond. Loss &  &   \\ \toprule
iUNet 
& Normal & Yes & $34.92 \pm 4.19$ & $0.810 \pm 0.148$ \\
&        & No  & $34.69 \pm 4.13$ & $0.806 \pm 0.151$ \\
\cmidrule{2-5}
& Radial & Yes & \colorbox{gray!20}{$34.99 \pm 4.39$} & $0.825 \pm 0.159$ \\
&        & No  & $34.75 \pm 4.23$ & $0.819 \pm 0.153$ \\
\bottomrule
\end{tabular}
\label{tab:lodopab_results}
\end{specialtable}

\begin{figure}
    \centering
    \includegraphics[width=\columnwidth]{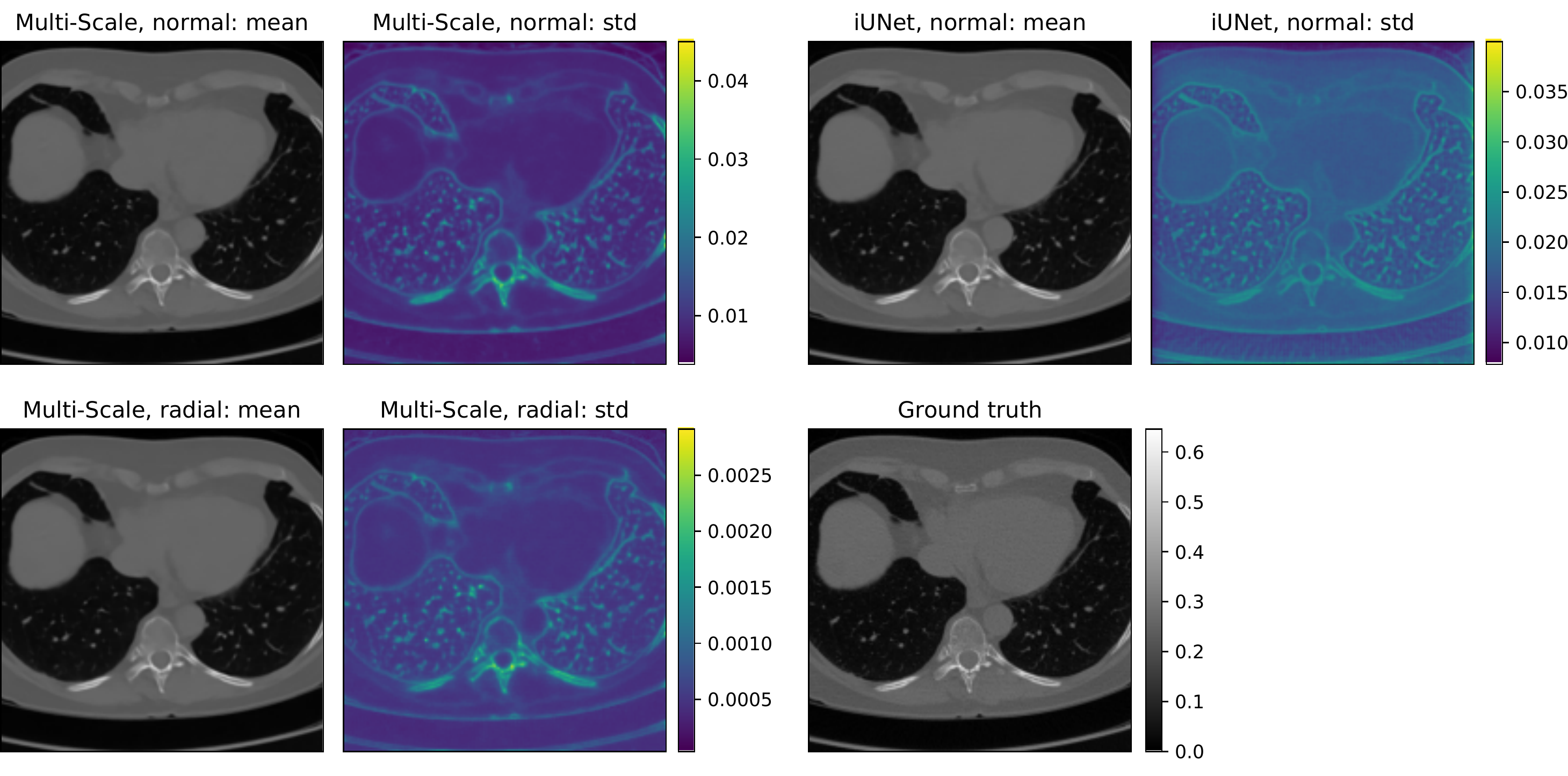}
    \caption{Cond.\ mean and point-wise standard deviation for the iUNet and the multi-scale architecture on the   \textit{LoDoPaB-CT} data.}
    \label{fig:lodopab_mean_std}
\end{figure}

Table \ref{tab:lodopab_results} shows mixed results for the different base distributions. While the iUNet benefits from the choice of the radial Gaussian distribution, the performance is worse for the multi-scale model. Nevertheless, the difference in PSNR and SSIM is only minor in this test. However, we could observe a difference in the quality and deviation during the sampling process for a single reconstruction. Networks that were trained with the radial distribution could produce high-quality reconstructions from a single sample. On the other hand, the standard deviation between each sampled reconstructions is significantly smaller than for the models with normal distribution. This can also be seen in the standard deviation plots in Figure \ref{fig:lodopab_mean_std}. Overall, models trained with the radial distribution can use fewer samples for the conditional mean to achieve good reconstructions. 

\subsubsection{Training with Additional Noise}
In most image datasets, pixel values can only take a specific, discrete range of values. Training a continuous flow-based model on discrete data can lead to artifacts, i.e.\ the model allocates arbitrary high likelihood values to the discrete values \citep{uria2013rnade}. In order to circumvent this problem, it is common to add a small amount of noise to the data to get a continuous distribution. This process is called dequantization and, in recent reviews, is done on all image datasets \citep{kobyzev2020normalizing}. We found that this problem was not as severe for the medical imaging datasets studied in this paper, e.g.\ the \textit{LoDoPaB-CT} dataset already used a dequantization of the discrete HU values. There is, however, a different problem with medical imaging datasets used for image reconstruction. Since there are no real ground truth images available, high-quality reconstructions are used for training. For \textit{LoDoPaB-CT}, reconstruction from normal-dose CT measurements and for \textit{fastMRI} reconstruction from fully sampled MRI measurements are used instead \citep{Leuschner_2021, zbontar2019fastmri}. These reconstructions are not free of noise, so we use an additional dequantization step and add random Gaussian noise in the order of the background noise to the training images. As an ablation, we add random Gaussian noise with zero-mean and a variance of $0.005$ to the ground truth images during training. We have chosen these values to correspond to the empirical background noise in the ground truth images. 

In Table \ref{tab:lodopab_results}, results for the multi-scale network and the iUNet with and without additional training noise are shown. For the multi-scale architecture, there is no visible difference between the two setups. The iUNet performs better without additional noise on the training images. In this case, the PSNR and SSIM values are very close to the best multi-scale network. Due to the high number of images in the dataset (lower overfitting risk) and the existing dequantization, additional noise does not seem to be beneficial in this case.

\subsubsection{Training with Conditional Loss}
As described in Section \ref{sec:iUnet}, the final output of the conditional network for the iUNet is in the image domain $X$. As an ablation, we added a supervised mean squared error loss to the negative log-likelihood term, see Equation~\eqref{eq:cond_loss_mse}, during the training using a weighting factor $\alpha = 1.0$. This additional loss could guide the conditional network to learn more relevant features. 

The results for the iUNet are given in the lower part of Table \ref{tab:lodopab_results}. The network benefits from the additional loss on the output of the conditioning network. However, like for all regularization terms, putting too much weight on the conditioning loss interferes with the primary objective of the cINN model. The performance deteriorates in this case. The loss also has a direct impact on the intermediate representations of the conditioning UNet. They shift from feature selection to the reproduction of complete reconstructions. An example is shown in Figure \ref{appfig:cond_and_no_cond} in the Appendix.

\subsubsection{Sample Refinement}
\label{sec:sample_refinement}
Using cINN, we are able to sample realistic-looking CT reconstruction. However, we have no guarantees that the sample explains the data $y$, i.e.\ $A T_\theta(y,z) \approx y$. In order to fulfill this data consistency constraint, we use an additional refinement based on a variational Tikohonov formulation: 
\begin{align}
\label{eq:sample_refinement}
    \hat{x} \in \arg \min_x \| A x - y \|_2^2 - \lambda \log p_\theta(x|y).
\end{align}
We solve for $\hat{x}$ using an iterative scheme and use as initialization our sample $T_\theta(y,z)$ from the cINN. In our experiments, only using the maximum posterior solution as a reconstruction often results in artifacts in the reconstructed image. Therefore, we transitioned to the penalized version in Equation \eqref{eq:sample_refinement}. An important topic is the choice of the parameter $\lambda$. In Table \ref{tab:sample_refinment}, the results for both the iUNet and the multi-scale architecture are given. Increasing the weighting factor $\lambda$ from $0$ to $1.0$ leads to an improvement in terms of PSNR and SSIM for both the multi-scale architecture and the iUNet. However, further increasing the factor $\lambda$ leads again to a deterioration in most cases. 

In total, the reconstruction quality with the sample refinement is worse than for the conditional mean approach. Therefore, we stick to the conditional mean reconstruction technique for the following experiments on the fastMRI dataset.

\begin{specialtable}
\caption{Sample Refinement for \textit{LoDoPaB} on the first $100$ samples of the test set. Minimized Equation \eqref{eq:sample_refinement} for $100$ iterations with a learning rate \num{1e-4}. Initial value was one sample from our model $x_0 = T_\theta^{-1}(z, y^\delta)$.}
\begin{tabular}{cccc}
\multicolumn{4}{c}{\textit{LoDoPaB-CT}} \\
\toprule Model & $\lambda$ & PSNR & SSIM  \\
\toprule Multi-scale & $0$ & $32.02 \pm 3.18$ & $0.742 \pm 0.135$  \\      
& $0.01$ & $32.10 \pm 3.21$ & $0.749 \pm 0.137$  \\
& $0.1$ & $32.56 \pm 3.40$ & $0.766 \pm 0.142$  \\
& $1.0$ &  \colorbox{gray!20}{$33.03 \pm 3.58$} & $0.783 \pm 0.148$  \\
& $10.0$ & $32.97 \pm 3.56$ & \colorbox{gray!20}{$0.784 \pm 0.149$}  \\
\midrule iUNet & $0$ & $32.16 \pm 3.12$ & $0.731 \pm 0.126$  \\   
& $0.01$ & $32.31 \pm 3.19$ & $0.737 \pm 0.128$  \\
& $0.1$ & $32.83 \pm 3.41$ & $0.759 \pm 0.135$  \\
& $1.0$ & \colorbox{gray!20}{$32.98 \pm 3.45$} & \colorbox{gray!20}{$0.765 \pm 0.136$}  \\
& $10.0$ & $32.88 \pm 3.40$ & $0.756 \pm 0.133$  \\
 \bottomrule \\
\end{tabular}
\label{tab:sample_refinment}
\end{specialtable}

\subsection{Magnetic Resonance Imaging}
\label{sec:results_mri}
The results for the two architectures, multi-scale and iUNet, for different configurations are presented in Table \ref{tab:fastmri_results}. Example reconstructions and point-wise standard deviations between samples for the best models are shown in Figure \ref{fig:fastmri_mean_std}. For all configurations, the models were trained using the Adam optimizer \cite{kingma2014adam}, and the initial learning rate of $\num{1e-4}$ was reduced by a factor of \num{0.8} on plateaus. The final model was chosen as the best model regarding the negative log-likelihood on the validation set. As the ground truth images for the \textit{fastMRI} test set are not publicly available, we report the PSNR and SSIM on the validation data in Table \ref{tab:fastmri_results}. Further, following the evaluation in \cite{zbontar2019fastmri}, we present the results subdivided into PD and PDFS.

Both networks were implemented such that the number of parameters was comparable ($2.5$ Mio.\ for the iUNet and $2.6$ Mio.\ for the multi-scale network). We used \num{5} scales for all networks. For the iUNet additive coupling layers and for the multi-scale architecture, affine coupling layers were used. The conditioning network for the iUNet was based on a UNet architecture. For the multi-scale network, we used an architecture based on a ResNet. Both use the zero-filled IFFT as model-based inversion layer.

\end{paracol}
\nointerlineskip
\begin{specialtable}
\caption{Results for the \textit{fastMRI} dataset. Conditioned mean computed with $100$ samples. Unless otherwise specified no additional training noise and no cond. loss were used.}
\begin{tabular}{ccccccc}
\multicolumn{7}{c}{\textit{fastMRI}} \\
\toprule Model & Base Distribution & Train Noise & \multicolumn{2}{c}{PSNR} & \multicolumn{2}{c}{SSIM}  \\
& & & PD & PDFS & PD & PDFS \\
\toprule Multi-scale & Normal & Yes & $29.15 \pm 6.25$ & $23.18 \pm 8.20$ & \colorbox{gray!20}{$0.777 \pm 0.086$} & \colorbox{gray!20}{$0.536 \pm 0.105$}  \\
 & & No & $28.54 \pm 6.52$ & $20.92 \pm 9.87$ & $0.776 \pm 0.086$ & \colorbox{gray!20}{$0.536 \pm 0.105$}  \\
 \cmidrule{2-7} & Radial & Yes & $31.84 \pm 3.56$ & $25.76 \pm 5.92$ & $0.760 \pm 0.092$ & $0.515 \pm 0.107$  \\
 & & No & \colorbox{gray!20}{$32.07 \pm 2.34$} & \colorbox{gray!20}{$26.54 \pm 2.73$} & $0.764 \pm 0.090$  & $0.522 \pm 0.103$  \\ \toprule
 iUNet & Normal & No & $27.85 \pm 1.38$ & $25.76 \pm 2.10$ & $0.622 \pm 0.052$ & $0.474 \pm 0.096$ \\
 \cmidrule{2-7} & Radial & No & $31.89 \pm 2.43$ & $25.94 \pm 2.86$ & $0.732 \pm 0.107$ & $0.432 \pm 0.126$ \\  \bottomrule \toprule & & Cond. Loss \\ \toprule 
iUNet & Normal & Yes & $27.91 \pm 1.35$ & $25.83 \pm 2.12$ & $0.628 \pm 0.054$ & $0.474 \pm 0.096$  \\
 & & No & $27.85 \pm 1.38$ & $25.76 \pm 2.10$ & $0.622 \pm 0.052$ & $0.474 \pm 0.096$   \\ 
 \cmidrule{2-7} & Radial & Yes & $31.62 \pm 2.26$ & $26.04 \pm 2.81$ & $0.730 \pm 0.096$ & $0.469 \pm 0.110$  \\
 & & No & $31.89 \pm 2.43$ & $25.94 \pm 2.86$ & $0.732 \pm 0.107$ & $0.432 \pm 0.126$  \\\bottomrule
\end{tabular}
\label{tab:fastmri_results}
\end{specialtable}
\begin{paracol}{2}
\makeatletter\ifthenelse{\equal{\@status}{submit}}{\linenumbers}{}\makeatother
\switchcolumn

\subsubsection{Base Distribution}
As with the LoDoPaB dataset, we investigate the influence of the target distribution. The results in Table \ref{tab:fastmri_results} show that switching from the standard Gaussian distribution to the radial Gaussian leads to an improvement in terms of PSNR for nearly all configurations on the fastMRI dataset. This is in contrast to the observations on the LoDoPaB dataset, where the differences were only minor. An explanation can be the lower number of samples used for the conditioned mean compared to the experiments on LoDoPaB (\num{100} vs.\ \num{1000}). The performance of models trained with the normal Gaussian base distribution highly depends on a sufficient number of samples for the reconstruction. On the other hand, an increase in the number of samples usually leads to an equivalent increase in the computing time. On fastMRI, we could also observe higher PSNR and SSIM values for single sample reconstruction from models with radial Gaussian base distribution.

\subsubsection{Training with Additional Noise}
We follow the same noising strategy as for the LoDoPaB-CT data and
add random Gaussian noise with zero-mean and a variance of $0.005$ to the ground truth images during training. For the multi-scale architecture, we observe an improvement for the standard Gaussian and a decline for the radial Gaussian base distributions. We noticed instabilities during the training for the iUNet. Therefore, only values without additional noise are given in Table \ref{tab:fastmri_results}.

\begin{figure}
    \centering
    \includegraphics[width=\columnwidth]{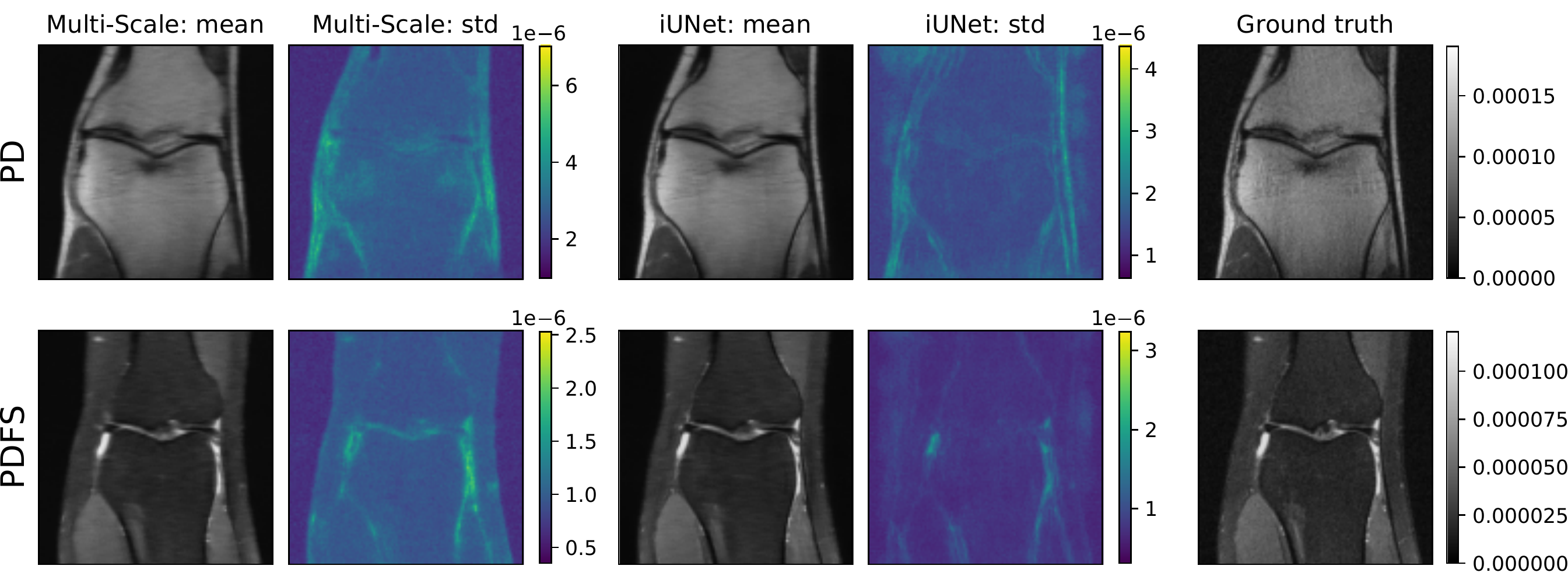}
    \caption{Cond.\ mean and point-wise standard deviation for the best-performing multi-scale architecture and iUNet on the   \textit{fastMRI} data. Both networks use the radial base distribution, no additional training noise, and the iUNet is trained with conditional loss. }
    \label{fig:fastmri_mean_std}
\end{figure}

\subsubsection{Training with Conditional Loss}
The results for the inclusion of the conditional loss term are given in the lower part of Table \ref{tab:fastmri_results}. On fastMRI, introducing this additional term to the loss function only gives a slight improvement in terms of PSNR and SSIM. In fact, we also observe a minor deterioration for the iUNet trained using the radial Gaussian base distribution on the PD case for \textit{fastMRI}.

\section{Discussion}
In this work, we have studied various configurations of conditioned flow-based models on different datasets. The focus of our research was to determine best practices for the use of cINN models for reconstruction tasks in CT and MRI. The two networks used, multi-scale and iUNet, showed comparable performance in many cases. The results demonstrate that a crucial part of the cINN models is the design of the conditioning network. A precise model-based inversion layer and a subsequent, extensive neural network can provide diverse features for the CNF. In particular, the model-based layer forms an interesting basis for combining mathematical modeling and data-driven learning. This can go much further than the FBP and Fourier models used here.

The choice of the base distribution also has a significant impact on the model's performance. The radial Gaussian proved to be a valuable alternative to the normal Gaussian distribution. Primarily to reduce the reconstruction time by needing fewer samples for the conditioned mean and avoiding common problems with high-dimensional distributions. For the noising during training and the additional conditioning loss, on the other hand, there is no clear recommendation. The additional noise might help on small datasets, where it acts as a data augmentation step. The conditioning loss requires extra tuning of the weighting factor. More promising, therefore, might be the use of a pre-trained reconstruction network whose parameters are frozen for use in cINN.

The experiments have also indicated that the training of cINN models does not always run without problems. Although invertible neural networks are analytically invertible, it is possible to encounter instabilities in some situations, and the networks may become numerically non-invertible. Furthermore, in this work, we have used the conditional mean as a reconstruction method for most of the experiments. However, other choices are possible. In the following, we will address these topics in more detail.

\subsection{Stability}
Recently it was noted that due to stability issues, an extensive invertible neural network could become numerically non-invertible in test time due to rounding errors \citep{behrmann2020understanding}. We observed this problem when evaluating iUNets with affine coupling layers. In Figure \ref{fig:iUnet_stability}, we show the loss during training and an example reconstruction after training. It can be observed that even when the training looks stable, one can get severe artifacts on unknown test images. We did not observe this problem for the multi-scale architecture. Affine coupling layers can have arbitrary large singular values in the inverse Jacobian matrix, which leads to an unstable inverse pass. This effect is known as \textit{exploding inverse} \citep{behrmann2020understanding}. For increasing stability in the iUNets, we suggest using additive coupling blocks in this architecture.  

Also, the inclusion of additional training noise led to severe instability in our experiments with the iUNet on the \textit{fastMRI} data. We did not obtain any meaningful reconstructions for this case. In contrast, these issues neither occurred with the multi-scale architecture on \textit{fastMRI} nor with the iUNet on \textit{LoDoPaB-CT}.

\begin{figure}[h]
    \centering
    \includegraphics[scale=0.475]{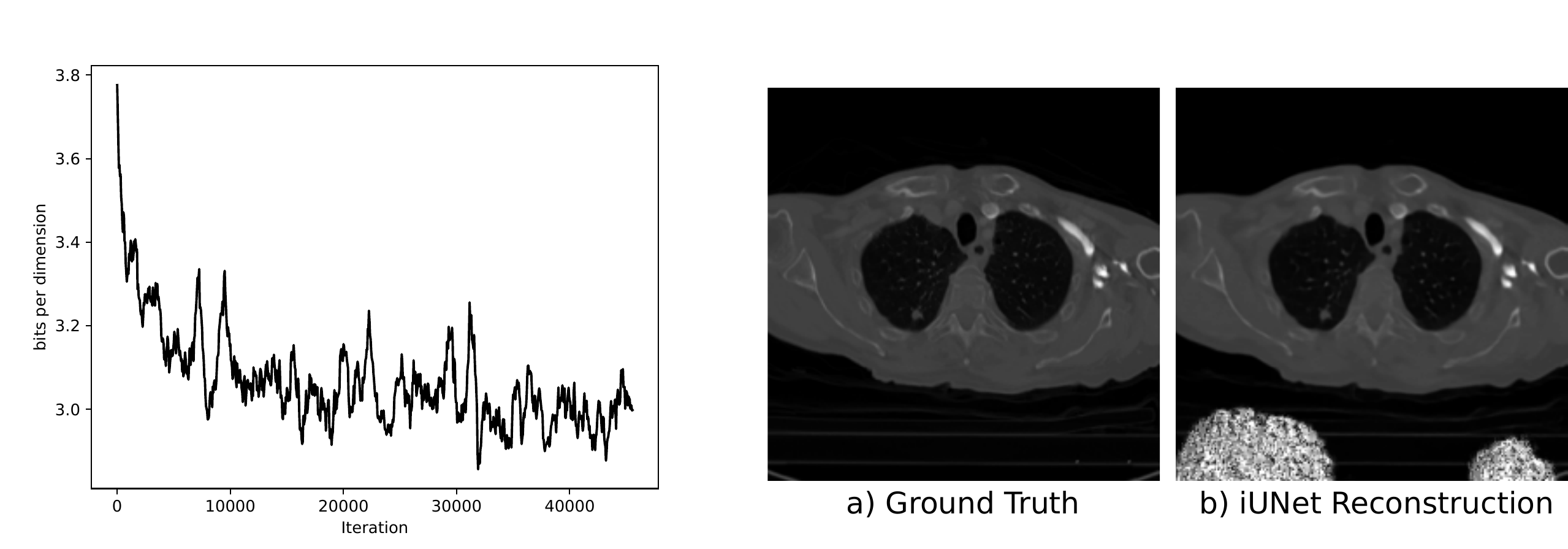}
    \caption{Left: Moving average of loss during Training. Right: Ground truth image from the \textit{LoDoPaB} test dataset and the corresponding iUNet reconstruction. The pixels in white visualize exploding values in the reconstruction.}
    \label{fig:iUnet_stability}
\end{figure}

\subsection{Reconstruction Method}
A trained cINN offers us the possibility to explore the full posterior. However, for evaluating the reconstruction quality of our models, we use the conditioned mean as a point estimate. This was also done in prior work for computed tomography reconstruction \citep{Denker2020, leuschner2021quantitative}, but it would be interesting to explore different choices of estimates. In Section \ref{sec:sample_refinement}, we evaluated a penalized version of the maximum posterior estimate. For the \textit{LoDoPaB-CT} dataset, this results in a lower PSNR and SSIM compared to using the conditioned mean. However, one could combine the idea of the conditioned mean and the sample refinement to combine samples that have a low, regularized data discrepancy (cf.\ Equation \eqref{eq:sample_refinement}).

\section{Conclusions}
This work has explored different architectures and best practices for applying conditional flow-based methods to medical image reconstruction problems in CT and MRI. Our experiments included two popular, invertible network designs. The iUNet \cite{etmann2020iUnets} architecture is inspired by the commonly used UNet \cite{ronneberger2015unet} used extensively in imaging applications. The multi-scale architecture is used in all major normalizing flow frameworks, like Glow \citep{kingma2018glow} or NICE \citep{dinh2015nice}. The invertible architectures were combined with a conditioning network, which extracts various features from the measurements for the reconstruction process. This cINN framework combines the advantages of memory-efficient invertible networks and normalizing flows for uncertainty estimation with a versatile reconstruction model. Additionally, it provides a direct way to combine model-based and data-driven approaches in a single model.   

The use of cINN models for medical image reconstruction is in its beginning stages, and many possible improvements should be explored. We investigated the radial Gaussian distribution as an alternative to the normal Gaussian base distribution. Our experiments show that it can be beneficial in many cases. A promising next direction is the development of novel invertible network architectures from existing approaches. For applications in medical image reconstruction, state-of-the-art deep learning methods are based on unrolled iterative methods \citep{adler2017solving}. In \citep{leuschner2021quantitative}, an extensive evaluation of the \textit{LoDoPaB-CT} dataset was performed, and the best scoring deep learning method was an unrolled learned primal-dual algorithm \citep{adler2018learned}. These unrolled iterative methods can be made invertible \citep{putzky2019invert} but are currently only used for memory-efficient backpropagation. In further work, we want to evaluate whether invertible iterative architectures can be integrated into flow-based models. 
\vspace{6pt} 



\authorcontributions{Conceptualization, A.D., M.S. and J.L.; Software, A.D., M.S. and J.L.; supervision, P.M.; project administration, P.M.; writing---original draft preparation, A.D., M.S. and J.L.; writing---review and editing, A.D., M.S. and J.L.; visualization, A.D.; funding acquisition, P.M. All authors reviewed, finalized and approved the manuscript.}


\funding{J.L., M.S., A.D. and P.M. were funded by the German Research Foundation (DFG; GRK 2224/1). J.L. and M.S. additionally acknowledge support by the project DELETO funded by the Federal Ministry of Education and Research (BMBF, project number 05M20LBB). A.D. further acknowledges support by the Klaus Tschira Stiftung via the project MALDISTAR (project number 00.010.2019).}

\conflictsofinterest{The authors declare no conflict of interest.}




\appendixtitles{no} 
\appendixstart
\appendix
\section{}
\subsection{Radial Density}
\label{app:radial_distr}
High-dimensional normal distributions do not behave intuitively as known from the low-dimensional settings. Sampling from a normal distribution gives mostly samples in the \textit{typical set}. For an $n$-dimensional normal distribution $\mathcal{N}(\mu, \sigma^2)$, this typical set has a distance of $\sigma \sqrt{n}$ from the expected value $\mu$. In \citep{farquhar2020radial}, the authors propose an $n$-dimensional radial Gaussian density in hyperspherical coordinates where
\begin{itemize}
    \item the radius $r$ is distributed accordingly to a half-normal distribution,  
    \item all angular components $\varphi_1, \dots, \varphi_{n-2} \in [0, \pi], \varphi_{n-1} \in [0, 2\pi]$ are uniformly distributed, yielding equal probability density at every point on the surface of the $n$-dimensional sphere. 
\end{itemize}
Our derivation of the likelihood closely follows \citep{farquhar2020radial}. We assume that all dimensions are independently distributed. For the radius $r$, we get the density:
\begin{align}
    p(r;\theta) = \frac{\sqrt{2}}{\sqrt{\pi} \sigma} \exp(-\frac{r^2}{2 \sigma^2}) \quad \text{ for } r \ge 0.
\end{align}
Let $v$ be a point on the unit sphere. We want every point on the unit sphere to be equally likely, i.e.\ 
\begin{align}
p(v) = \frac{1}{S_n} \quad \text{ with } S_n = 2\frac{\pi^{n/2}}{\Gamma(n/2)},    
\end{align}
where $S_n$ is the surface of the $n$-dimensional unit sphere. We can get the density for the radial components $p(\phi_1, \dots, \phi_{n-1})$ by solving
\begin{align}
\label{eq:radial_density_1}
    p(v) \dd A = \frac{1}{S_n} \dd A = p(\phi_1, \dots, \phi_{n-1}) \dd \phi_1 \dots \dd \phi_{n-1}. 
\end{align}
Here, $\dd A$ is the surface element:
\begin{align}
    \dd A =  \dd{\phi_{n-1}} \prod_{i=1}^{n-2} \sin(\phi_i)^{n-1-i} \dd{\phi_1} \dots \dd{\phi_{n-2}}.
\end{align}
Solving \eqref{eq:radial_density_1} leads us to the density:
\begin{align}
    p(\phi_1, \phi_2, \dots, \phi_{n-1}) = \frac{1}{S_n} \prod_{i=1}^{n-2} \sin(\phi_i)^{n-1-i}.
\end{align}
Setting $\sigma = 1$ for the radial components gives us the full density in hyperspherical coordinates: 
\begin{align}
\label{eq:DensitySpherical}
    p_\epsilon(\epsilon = (r, \phi_1, \phi_2, \dots, \phi_{n-1})) = \frac{\sqrt{2}}{\sqrt{\pi}} \exp(- r^2/2) \frac{1}{S_n} \prod_{i=1}^{n-2} \sin(\phi_i)^{n-1-i}.
\end{align}
For our experiments, we are always working in Cartesian coordinates, so one has to do a final transformation $x = f(\epsilon)$ and use the change-of-variables theorem. The Jacobian of the $n$-dimensional spherical coordinate transformation is known: 
\begin{align}
    \left|\det\left(\frac{\partial f(\epsilon)}{\partial \epsilon}\right)\right| = r^{n-1} \prod_{i=1}^{n-2} \sin(\phi_i)^{n-1-i}.
\end{align}
Finally, we get 
\begin{equation}
\begin{aligned}    
p_x(x) &= \frac{\sqrt{2}}{ \sqrt{ \pi} S_n} \exp(-r^2/2) \prod_{i=1}^{n-2} \sin(\phi_i)^{n-1-i} \left(r^{n-1} \prod_{i=1}^{n-2} \sin(\phi_i)^{n-1-i}\right)^{-1} \\ 
    &= \frac{\sqrt{2}}{\sqrt{ \pi} S_n r^{n-1}} \exp(-r^2/2) \\ 
    &= \frac{\sqrt{2}}{\sqrt{ \pi} S_n \|x\|^{n-1}} \exp(-\|x\|^2/2)
\end{aligned}
\end{equation} 
as our radial Gaussian density.

\subsection{Architecture for Compressed Sensing Example}
\label{app:comp_sense_arch}
The multi-scale architecture used for the compressed sensing experiments used in Section \ref{sec:compressed_sensing} consists of two learnable downsampling operations, each followed by a conditional coupling block. After a flatten layer, a last dense conditional coupling block is used.

\begin{tabular}{|l|l|}
\hline cINN & Output size \\
\hline Learnable Downsampling & $4 \times 14 \times 14$ \\
\hline level 1 conditional section &  $4 \times 14 \times 14$   \\ 
\hline Learnable Downsampling & $16 \times 7 \times 7$ \\
\hline level 2 conditional section &  $16 \times 7 \times 7$   \\ 
\hline Flatten & $784$ \\
\hline Split: $656$ to output & $128$ \\
\hline level 3 Dense-conditional section & $128$ \\
 \hline 
\end{tabular}{}

In the conditional coupling section, we use an affine coupling layer and implemented the scale $s$ and translation $t$ using a small convolutional neural network with either $1\times 1$ convolution or $3 \times 3$ convolutions. After each affine coupling layer, we use a fixed $1 \times 1$ convolution to permute the dimensions. For the dense coupling section, we use a simple random permutation of the dimensions and affine coupling layers with dense subnetworks $s$ and $t$.

\begin{tabular}{|l|l|}
\hline
\multicolumn{2}{|l|}{conditional section} \\ \hline
 Affine coupling (CNN-subnet with $1 \times 1$ kernel) & \multirow{4}{*}{8x}          \\ \cline{1-1}
 $1 \times 1$ convolution &   \\ \cline{1-1}
 Affine coupling (CNN-subnet with $3 \times 3$ kernel)  &   \\ \cline{1-1}
 $1 \times 1$ convolution & \\ \hline
\end{tabular}

\begin{tabular}{|l|l|}
\hline
\multicolumn{2}{|l|}{dense conditional section} \\ \hline
   Random permutation & \multirow{2}{*}{3x}             \\ \cline{1-1}
   Affine coupling (Dense-subnetwork) &   \\ \hline
\end{tabular}

\subsection{Evaluation Metrics}
\label{app:eval_metrics}
\subsubsection{Peak Signal-to-Noise Ratio}
The peak signal-to-noise ratio (PSNR) is measured by a log-scaled version of the mean squared error (MSE) between the reconstruction $\hat{x}$ and the ground truth image $x^\dagger$. PSNR expresses the ratio between the maximum possible image intensity and the distorting noise
\begin{align*}
    \operatorname{PSNR}\left(\hat{x}, x^\dagger\right) := 10 \log_{10} \left( \frac{L^2}{\operatorname{MSE}\left(\hat{x}, x^\dagger \right)} \right), \quad \operatorname{MSE}\left(\hat{x}, x^\dagger \right) := \frac{1}{n} \sum_{i=1}^{n} \left\vert \hat{x}_i - x^\dagger_i \right\vert^2
\end{align*}

In general, higher PSNR values are an indication of a better reconstruction. The maximum image value $L$ can be chosen in different ways. For the MNIST and the LoDoPab-CT dataset, we compute the value per slice as $L=\max(x^\dagger) - \min(x^\dagger)$. For evaluation on the fastMRI dataset, we choose $L$ as the maximum value per patient, i.e.\ per $3$D volume.


\subsubsection{Structural Similarity}
The structural similarity (SSIM) \cite{wang2004image} compares the overall image structure of ground truth and reconstruction. It is based on assumptions about human visual perception. Results lie in the range $[0,1]$, with higher values being better. The SSIM is computed through a sliding window at $M$ locations
\begin{align*}
    \operatorname{SSIM}\left(\hat{x}, x^\dagger\right) := \frac{1}{M} \sum_{j=1}^{M} \frac{\left(2 \hat{\mu}_j \mu_j + C_1 \right) \left(2 \Sigma_j + C_2 \right) }{\left( \hat{\mu}_j^2 + \mu_j^2 + C_1 \right) \left(\hat{\sigma}_j^2 + \sigma_j^2 + C_2\right)}.
\end{align*}

Here, $\hat{\mu}_j$ and $\mu_j$ are the average pixel intensities, $\hat{\sigma}_j$ and $\sigma_j$ the variances and $\Sigma_j$ the covariance of $\hat{x}$ and $x^\dagger$ at the $j$-th local window. Constants $C_1 = (K_1 L)^2$ and $C_2 = (K_2 L)^2$ stabilize the division. Just as with the PSNR metric, the maximum image value $L$ can be chosen in different ways.
We use the same choices as specified in the previous section.

\subsection{Additional Figures}
\begin{figure}[h!]
    \centering
    \includegraphics[width=0.65\textwidth]{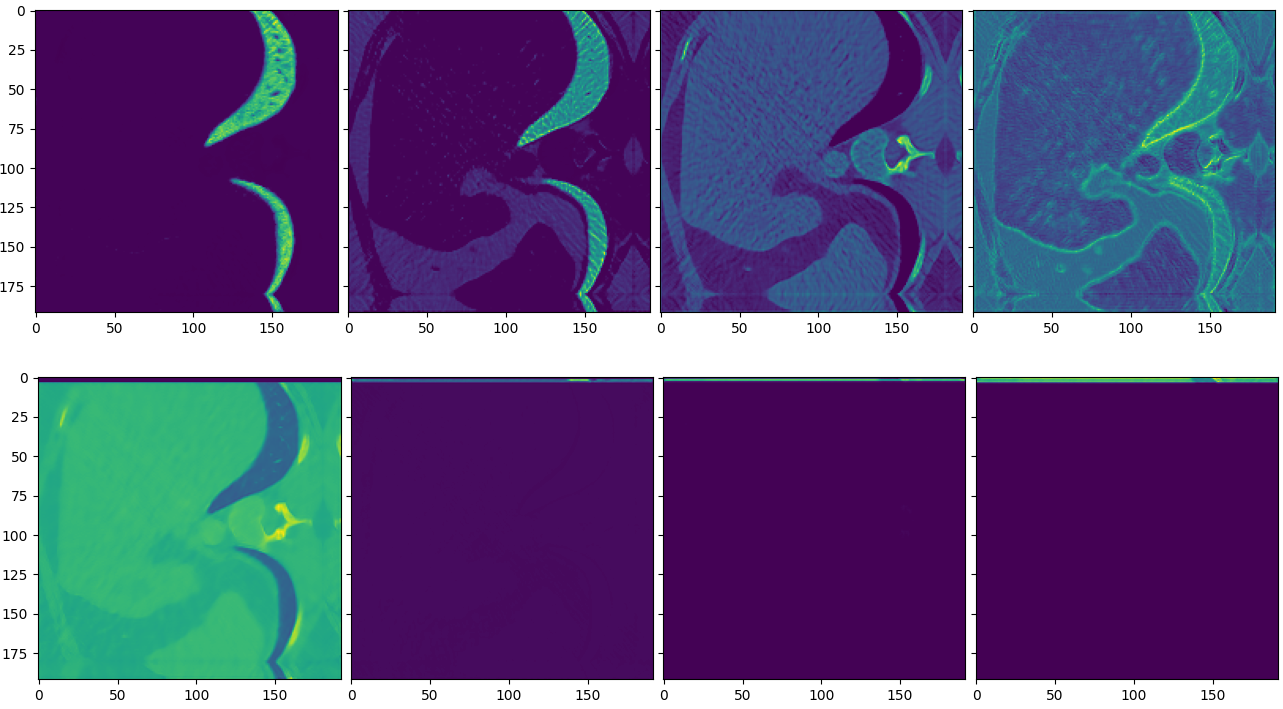}
    \caption{Intermediate activations from a single layer in a conditioning UNet model. Top: cINN Model trained with just the log-likelihood. Bottom: cINN model trained with an additional loss for the conditioning UNet. One can observe, that the conditioning network focuses on different parts of the image, if no special loss is used. Otherwise, it produces activations that are close to the final reconstruction. In addition, there are many empty activations.}
    \label{appfig:cond_and_no_cond}
\end{figure}

\end{paracol}  
\reftitle{References}


\externalbibliography{yes}
\bibliography{references.bib}

%


\end{document}